\documentclass[prd,nofootinbib,notitlepage]{revtex4-1}
\usepackage{epsfig,amsmath,amssymb,slashed}
\usepackage{graphicx}
\usepackage{hyperref}
\usepackage{color}
\newcommand{\bra}[1]{\langle #1|}
\newcommand{\ket}[1]{|#1\rangle}

\newcommand{\di}{{\rm d}}

\newcommand{\Pro}{{\sf P}}

\def\wT{{\widehat T}}
\def\wj{{\widehat j}}

\def\wpsi{{\widehat{\psi}}}
\def\wrho{{\widehat{\rho}}}
\def\wrhol{{\widehat{\rho}_{\rm LE}}}

\newcommand{\tr}{{\rm tr}}  
  
\newcommand{\e}{{\rm e}}

\newcommand{\piv}{\mbox{\boldmath$\pi$}}
\newcommand{\omegav}{\boldsymbol{\omega}}

\newcommand{\x}{{\rm x}}

\newcommand{\be}{\begin{equation}}
\newcommand{\ee}{\end{equation}}                                                                               
\newcommand{\bea}{\begin{eqnarray}}
\newcommand{\eea}{\end{eqnarray}}                                                                               
\newcommand{\bn}{n_B}

\begin{document}

\begin{center}
\Large\bf{Local thermodynamical equilibrium\\ and the $\beta$ frame for a
quantum relativistic fluid}
\end{center} 
\medskip
\begin{center} 
Francesco Becattini \footnote{e-mail: becattini@fi.infn.it} \\
\vspace{0.3 cm}
{\it Universit\`a di Firenze and INFN Sezione di Firenze, Florence, Italy}\\
\vspace{0.3 cm}
Leda Bucciantini \footnote{e-mail: leda.bucciantini@df.unipi.it} \\
\vspace{0.3 cm}
{\it Dipartimento di Fisica dell'Universit\`a di Pisa and INFN, 56127 Pisa, Italy}\\
\vspace{0.3 cm}
Eduardo Grossi \footnote{e-mail: grossi@fi.infn.it} \\
\vspace{0.3 cm}
{\it Universit\`a di Firenze and INFN Sezione di Firenze, Florence, Italy}\\
\vspace{0.3 cm}
Leonardo Tinti \footnote{e-mail: dr.leonardo.tinti@gmail.com}\\
{\it Jan Kochanowski University, Kielce, Poland}
\end{center}

\begin{abstract}
We discuss the concept of local thermodynamical equilibrium in relativistic
hydrodynamics in flat spacetime in a quantum statistical framework without 
an underlying kinetic description, suitable for strongly interacting fluids. We show that the
appropriate definition of local equilibrium naturally leads to the introduction 
of a relativistic hydrodynamical frame in which the four-velocity vector is the 
one of a relativistic thermometer at equilibrium with the fluid, parallel to
the inverse temperature four-vector $\beta$, which then becomes a primary quantity. 
We show that this frame is the most appropriate for the expansion of stress-energy 
tensor from local thermodynamical equilibrium and that therein the local laws 
of thermodynamics take on their simplest form. We discuss the difference between 
the $\beta$ frame and Landau frame and present an instance where they differ. 
\end{abstract}

\maketitle

\section{Introduction}
\label{intro}

In recent years, relativistic hydrodynamics has drawn much attention. Part of the 
revived interest \cite{Baier:2007ix,Kovtun:2004de} is owing to the successfull 
hydrodynamic description of the Quark Gluon Plasma formed in collisions of nuclei 
at very high energy \cite{Kolb:2001qz,Ollitrault:2008zz,Heinz:2013th,Baier:2006um,
Huovinen:2006jp,csernai,florkowski,cracow}. It is also known that hydrodynamics can be applied 
to a large portion of the phase diagram  of condensed matter systems presenting 
quantum critical points \cite{Damle,Hartnoll,Kovtun,Fritz}. Focusing on the Quark 
Gluon Plasma, close to the QCD critical temperature, the system is made of strongly
interacting quantum fields and does not apparently allow a description in terms 
of weakly-interacting quasiparticles \cite{schaefer,rajagopal}. Thus, the use of 
kinetic theory to describe it can be questioned, and yet, because the microscopic 
interaction length is small compared to its overall size, the system is actually 
a fluid.

In principle, hydrodynamics does not need an underlying kinetic theory nor a discrete 
particle substratum, even if its use can be very effective to obtain useful relations 
\cite{denicol}. Hydrodynamic is, in essence, the continuity equation of the {\em mean} 
value of the stress-energy tensor (and charge current) operator, which, being primarily
expressed in terms of quantum fields, does not need a single-particle distribution 
function $f(x,p)$. In fact, its momentum-space integral expression in terms of $f(x,p)$
can be obtained under special conditions, those which make kinetic approach suitable 
\cite{degroot}.

Consequently, all basic concepts in hydrodynamics should be defined independently of 
kinetic theory and of the single-particle distribution function. Indeed, 
while flow velocity is usually defined as an eigenvector of some current, like in
the Landau and Eckart's frames, another very basic notion of hydrodynamics, that is
local thermodynamic equilibrium (LTE), is defined in most textbooks by means of kinetic 
theory; for instance, by making the collisional integral of the (relativistic) 
Boltzmann equation vanishing \cite{degroot}. In fact, we will show in this work that
this is not the most general definition; in quantum statistical mechanics LTE can be 
defined as a maximum of the entropy with specific constraints \cite{balian}. 
Furthermore it will be shown that, in the relativistic context, such a definition 
naturally leads to the introduction of a four-vector field - the inverse temperature
four-vector $\beta$, which functions as a hydrodynamical velocity, giving rise to a 
new hydrodynamical frame other than the known Landau's and Eckart's.

Thus far, this four-vector field has been mostly considered as a secondary quantity, 
formed by multiplying the invariant temperature $1/T$ by an otherwise defined velocity 
four-vector $u$. Recently, Van and Biro \cite{van} have argued that $\beta$ in fact
defines a new independent frame, a conclusion that we fully support. Indeed, in this paper, 
we will reinforce it and demonstrate it in the most general quantum relativistic 
framework without resorting to kinetic arguments. We will show that it is much more 
natural and convenient to take $\beta$ as a primordial field related to the concept 
of LTE, so that the four-velocity of a relativistic fluid can be defined starting 
from the $\beta$ field and not vice-versa:
$$
       u(x) \equiv \frac{\beta}{\sqrt{\beta^2}}
$$

The paper is organized as follows. In sect.~\ref{local}, \ref{local2} we review the 
concept of local thermodynamical equilibrium in relativistic quantum statistical 
mechanics and introduce the $\beta$ frame. In sect.~\ref{thermometer} we will show 
how to operationally define the $\beta$ vector 
through an ideal relativistic thermometer, providing a better insight of its physical
meaning. In sect.~\ref{setensor} we will discuss the form of the stress-energy tensor 
in the $\beta$ frame, in sect.~\ref{frames} we will point out the difference between 
$\beta$ and Landau frames; finally in sect.~\ref{relhyd} we will discuss the separation 
between the ideal and dissipative part of the stress-energy tensor.

\subsection*{Notation}

In this paper we use the natural units, with $\hbar=c=K=1$.\\ 
The Minkowskian metric tensor is ${\rm diag}(1,-1,-1,-1)$; for the Levi-Civita
symbol we use the convention $\epsilon^{0123}=1$.\\ 
We will use the relativistic notation with repeated indices assumed to 
be saturated, however contractions of indices will be sometimes denoted with 
dots, e.g. $ u \cdot T \cdot u \equiv u_\mu T^{\mu\nu} u_\nu$. Operators in 
Hilbert space will be denoted by a large upper hat, e.g. $\widehat {\sf R}$ 
while unit vectors with a small upper hat, e.g. $\hat v$. We will work with a 
symmetric stress-energy tensor with an associated vanishing spin tensor.

\section{Local thermodynamical equilibrium in relativistic quantum statistical
mechanics}
\label{local}

In the most general framework of quantum statistical mechanics, LTE is defined 
by the maximization of the Von Neumann entropy $S= -\tr (\wrho \, \log \wrho)$, $\wrho$ 
being the density operator, with the constraints of fixed densities of energy, momentum 
and charge \cite{balian}. As has been mentioned in the Introduction, such a definition 
does not require any underlying kinetic theory; the only requirement is that densities 
significantly vary over distances much larger than the typical microscopic scale.

In non-relativistic thermodynamics, the LTE definition is an unambiguous one and 
leads to a unique density operator obtained by maximizing the function of $\wrho$, 
with $\tr \wrho = 1$, for any time $t$
\be\label{free}
   -\tr (\wrho \, \log \wrho) + \int \di^3 \x \; b({\bf x},t) 
   \left[ \langle {\widehat h} ({\bf x},t) \rangle - h({\bf x},t) \right] - 
   b({\bf x},t) {\bf v}({\bf x},t) \cdot \left[ \langle \widehat{\piv}({\bf x},t) 
   \rangle - \piv ({\bf x},t) \right] - \xi({\bf x},t)
   \left[ \langle {\widehat q} ({\bf x},t) \rangle - q({\bf x},t) \right],
\ee
where $h$, $\piv$ and $q$ are the actual values of the energy, momentum and particle 
(or charge) density respectively; $b = 1/T$ and $\xi = \mu/T$ are point-dependent 
Lagrange multiplier, as well as ${\bf v}$ whose meaning is the mean velocity 
of the particles. The symbol $\langle \rangle$ stands for the renormalized mean value
of the operators:
$$
  \langle \widehat A \rangle = \tr (\wrho \, \widehat A)_{\rm ren} .
$$
For the simple case of a free quantum field theory this corresponds to the normal 
ordering of creation and destruction operator and if $\widehat A$ is quadratic in
the fields to the subtraction of its vacuum expectation value:
$$
  \tr (\wrho \, \widehat A)_{\rm ren} = \tr (\wrho \, : \! \widehat A \!:) 
  = \tr (\wrho \, \widehat A ) - \bra{0} \widehat A \ket{0}.
$$

The density operator $\wrhol$ resulting from the maximization of (\ref{free}) is 
called LTE density operator:
\be\label{rhol} 
  \wrhol = \frac{1}{Z_{\rm LE}} 
  \exp \left[ -\int \di^3 \x \; b({\bf x},t) \widehat h ({\bf x},t) + 
   b({\bf x},t) {\bf v}({\bf x},t)\cdot \widehat{\piv} ({\bf x},t) - 
   \xi({\bf x},t) {\widehat q}({\bf x},t) \right],
\ee  
where $Z_{\rm LE}$ is the normalizing factor making $\tr \wrhol = 1$. The values 
of the Lagrange multipliers $b$, $\bf{v}$ and  $\xi$ are obtained enforcing 
$\langle \hat A \rangle=A$, where $A$ is respectively the actual value of the 
energy, momentum and charge of the system. A galileian transformation does not change 
the resulting density operator except for a shift of the parameter ${\bf v}$, but 
the entropy $S= -\tr(\wrhol \log \wrhol)$ is invariant. It should be emphasized
that $\wrhol$ is {\em not} the true density operator. Indeed, in the Heisenberg
picture the LTE density operator $\wrhol$ in eq.~(\ref{rhol}) is explicitely dependent 
on time through the time dependence of the operators, while the true density operator 
must be time-independent in the Heisenberg picture. The relation between the true 
density operator $\wrho$ and $\wrhol$ will be discussed in sect.~\ref{setensor}.
\begin{center}
\begin{figure}[ht]
\includegraphics[width=0.45\textwidth]{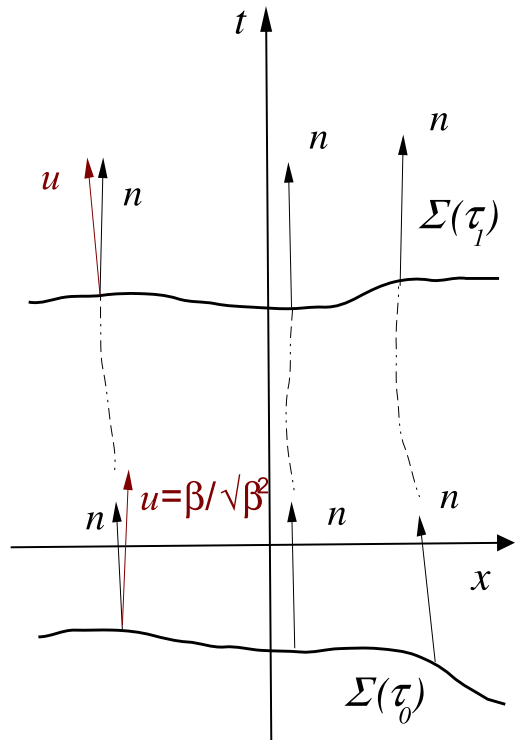}
\caption{Spacelike hypersurfaces $\Sigma(\tau)$ and their normal versor $n$ defining 
local thermodynamical equilibrium for a relativistic fluid in Minkwoski spacetime. If 
the $\beta$ field has vanishing vorticity, $\beta$ can be chosen  parallel to
$n$ at each point; conversely, the normal versor and the $\beta$ direction do not 
coincide (see sect.~\ref{local2}). 
\label{sigmas}}
\end{figure}
\end{center} 

Extending the definition of LTE to quantum relativistic statistical mechanics is not  
straightforward because energy density and momentum density are frame-dependent 
quantities in a much stronger fashion than in non-relativistic mechanics. To make 
it fully covariant, it is necessary to fix a $\tau$-parametric family of spacelike 
hypersurfaces $\Sigma(\tau)$. The timelike unit vector field $n(x)$ normal to the 
surfaces defines world lines of observers (see fig.~\ref{sigmas}), yet the parameter
$\tau$, in general, does not coincide with the proper time of comoving clocks. 
As it is known, for orthogonal surfaces to exist, the field $n(x)$ must be vorticity-free, 
i.e. it ought to fulfill the equation:
\be\label{pfaff}
   \epsilon_{\mu\nu\rho\sigma} n^\nu (\partial^\rho n^\sigma - \partial^\sigma 
   n^\rho) = 0.
\ee
For the comoving frame having $n$ as time direction, we enforce the mean energy-momentum 
and charge density to be the actual ones everywhere:
\bea\label{constr}
 && n_\mu \tr(\wrhol \, \wT^{\mu\nu}(x))_{\rm ren} = n_\mu
  \langle \wT^{\mu\nu}(x) \rangle_{\rm LE} \equiv n_\mu T^{\mu\nu}_{\rm LE}(x) 
  = n_\mu T^{\mu\nu}(x) \nonumber \\
 && n_\mu \tr (\wrhol \wj^{\mu}(x))_{\rm ren} = n_\mu
  \langle \wj^{\mu}(x) \rangle_{\rm LE} \equiv n_\mu j^{\mu}_{\rm LE}(x) = 
  n_\mu j^{\mu}(x)
\eea 
where $\wT$ is the stress-energy tensor operator and $\wj$ the conserved current 
(if any). The function to be maximized as a function of $\wrho$, with $\tr\wrho=1$,
at any $\tau$, reads:
$$
   -\tr (\wrho \log \wrho) + \int_{\Sigma(\tau)} \di \Sigma \; n_\mu 
   \left[ \left( \langle \wT^{\mu\nu}(x) \rangle - T^{\mu\nu}(x) \right) 
   \beta_\nu (x) - \left( \langle \wj^\mu(x) \rangle - j^\mu(x) \right) 
   \xi(x) \right] 
$$
where $\di \Sigma$ is the measure (in the Minkowski spacetime) of the hypersurface, 
$\beta$ is, by definition, the inverse temperature four-vector and $\xi$ a scalar 
field of Lagrange multipliers whose meaning will be clear shortly. The solution is:
\be\label{rhol2} 
  \wrhol = \frac{1}{Z_{\rm LE}} 
  \exp \left[ -\int_{\Sigma(\tau)} \di \Sigma \; n_\mu \left( \wT^{\mu\nu}(x) 
  \beta_\nu(x) - \xi(x) \wj^\mu(x) \right) \right]
\ee  

This covariant form of an equilibrium density operator, to our knowledge, was first
obtained with this variational method by Zubarev \cite{zubarev}. It is clear that 
the operator in (\ref{rhol2}) does depend on the particular hypersurface $\Sigma(\tau)$ 
(whence on the field $n$). Accordingly, the mean values $T_{\rm LE}$ and $j_{\rm LE}$ 
depend on the hypersurface $\Sigma$, hence, in general one can write
$$ 
  T^{\mu\nu}_{\rm LE} = T^{\mu\nu}_{\rm LE}[\beta,\xi,n] \qquad  
  j^\mu_{\rm LE} = j^\mu_{\rm LE}[\beta,\xi,n],
$$
so that even the $\beta$ field, obtained as a solution of the eq.~(\ref{constr}), 
will depend on $n$:
\be\label{betaeq}
 n_\mu T^{\mu\nu}_{\rm LE} [\beta,\xi,n] = n_\mu T^{\mu\nu} \qquad
 n_\mu j^{\mu}_{\rm LE} [\beta,\xi,n] = n_\mu j^{\mu},
\ee
where the square brackets mean that the dependence of the currents on the fields
$\beta,\xi$ and $n$ is in general functional (e.g. there could be a dependence on
the derivatives). For all means to be independent of it, the divergence of the integrand 
should vanish - provided that some boundary conditions are enforced \cite{becacov} 
- a condition which is met if:
\be\label{killing}
  \partial_\mu \beta_\nu + \partial_\nu \beta_\mu = 0 
 \qquad \qquad \partial_\mu \xi = 0.
\ee
whose solution is (see also ref.~\cite{becacov}): 
\be\label{glob}
  \beta_\nu = b_\nu + \varpi_{\nu\lambda} x^\lambda \qquad \qquad \xi={\rm const}
\ee 
whence:
\be\label{glob2}
  \varpi_{\nu\lambda} = -\frac{1}{2} (\partial_\nu\beta_\lambda - \partial_\lambda
  \beta_\nu)
\ee
The above equations just define the known condition of global thermodynamical 
equilibrium (GTE) for a relativistic fluid: $\beta$ must be a Killing vector
\cite{chrobok}. 
They ensure the stationarity of the density operator, which now reads (from 
(\ref{rhol2})):
\be\label{globrho}
  \wrho = \frac{1}{Z} \exp \left[ -b_\nu \widehat P^\nu + \frac{1}{2} 
  \varpi_{\lambda\nu} \widehat J^{\lambda\nu} + \xi \widehat Q \right] = 
  \frac{1}{Z} \exp \left[ -\beta_\nu(x) \widehat P^\nu + \frac{1}{2} 
  \varpi_{\lambda\nu} \widehat J_x^{\lambda\nu} + \xi \widehat Q \right]
\ee 
where, in the rightmost expression, we have taken advantage of the translated angular 
momentum operator:
\be\label{angmomshift}
   \widehat J^{\lambda \nu} = \widehat J^{\lambda\nu}_x + x^\lambda \widehat P^\nu 
   - x^\nu \widehat P^\lambda,
\ee
The general GTE form (\ref{globrho}) depends - besides the chemical potential - on 
10 constant parameters, as many as the generators of the Poincar\'e group. The density
operator (\ref{globrho}) comprises all kwown instance of GTE in Minkowski spacetime
including the rotating equilibrium which will be further discussed in 
sect.~(\ref{frames}).

Going back to local equilibrium, as long as the field $n(x)$ is not specified, the 
definition of LTE is ambiguous\footnote{Note that the field $n$ does 
not necessarily coincide with the hydrodynamic velocity field $u$ albeit, as
we will see, it is related to it.}. To show that there is a preferential choice
thereof, one can calculate the total entropy by using (\ref{rhol})
\be\label{entropy}
 S = -\tr (\wrhol \log \wrhol)_{\rm ren} = \log Z_{\rm LE} + 
\int_{\Sigma(\tau)}  \di \Sigma \; n_\mu \left( T^{\mu\nu}_{\rm LE} \beta_\nu - \xi j^\mu_{\rm LE} 
 \right).
\ee
A crucial and mostly unspoken assumption in relativistic extension of thermodynamics
is that $\log Z_{\rm LE}$ can be written as an integral over the hypersurface $\Sigma$ of 
a four-vector field, defined as {\em thermodynamical potential current} $\phi^\mu$,
depending on the functions $\beta$ and $\xi$
\be\label{tpcurrent}
 \log Z_{\rm LE}(\tau) = \log \tr \left( \exp \left[ -\int_{\Sigma(\tau)}  \di \Sigma
 \; n_\mu \left( \wT^{\mu\nu} \beta_\nu - \xi \wj^\mu \right) \right] \right) 
 = \int_{\Sigma(\tau)}   \di \Sigma \; n_\mu \phi^\mu[\beta,\xi,n].
\ee
This assumption is necessary for the existence of an entropy current $s^\mu$ which is one 
of the starting points of Israel's formulation of relativistic hydrodynamics \cite{israel}. 
Although (\ref{tpcurrent}) should be proved, we account it in this work as an 
{\em ansatz}. Hence, in view of (\ref{entropy}) and (\ref{tpcurrent}), the entropy 
current reads:
\be\label{current} 
  s^\mu = \phi^\mu + T^{\mu\nu}_{\rm LE} \beta_\nu - \xi j^\mu_{\rm LE} + s^\mu_T(n)  
\ee
where $s_T(n)$ is an arbitrary four-vector field orthogonal to $n$. Note from 
(\ref{tpcurrent}) that also $\phi$ is defined up to an arbitrary four-vector field 
orthogonal to $n$. It should be emphasized that in nonequilibrium situations, since 
$\partial_\mu s^\mu \ne 0$, the total entropy $S$ in (\ref{entropy}) is a frame-dependent 
quantity, as it varies with the integration hypersurface $\Sigma$. Indeed, like 
$T_{\rm LE}$ and $j_{\rm LE}$, the local current $\phi$ will also depend on the 
hypersurface $\Sigma$ (see its dependence on $n$ in eq.~\ref{tpcurrent}).

A great simplification would be achieved if $n= \hat \beta = \beta_\mu/\sqrt{\beta^2}$,
as the number of independent variables on which mean values depend would be reduced.
With this choice the eq.~(\ref{constr}) would become
\be\label{betaeq2}
 \beta_\mu T^{\mu\nu}_{\rm LE} [\beta,\xi] = \beta_\mu T^{\mu\nu} \qquad
 \beta_\mu j^{\mu}_{\rm LE} [\beta,\xi] = \beta_\mu j^{\mu},
\ee
where the right hand sides contain the true values at each point. This choice is
what we define as $\beta$ {\em frame}, with a fluid velocity defined as:
$$
       u(x) \equiv \frac{\beta}{\sqrt{\beta^2}} \; .
$$

Indeed, setting $n = \hat \beta$ is possible only if the $\beta$ field solution of
the eq.~(\ref{betaeq}), also fulfills the equations (\ref{pfaff}). In fact, this 
equation does not apply even for the simple case of a rigid velocity field, which 
is actually a global thermodynamical equilibrium one (see Appendix A). Notwithstanding, 
also in the vorticous case, it is possible to find a proper definition of the $n$ 
field based on $\beta$, as it will be shown in the next section. 

From a physical viewpoint, the $\beta$ frame is identified by the four-velocity of a 
relativistic thermometer at local equilibrium with the system, what will be discussed 
in detail in sect.~\ref{thermometer}. This frame has more peculiar features. As an 
example, let us contract the equation (\ref{current}) with $n_\mu$, which enables us to 
use eq.~(\ref{constr}) to replace the local equilibrium averages  $T$ and $j$ with 
their actual values
\be\label{smunmu}
  s^\mu n_\mu = n_\mu \phi^\mu + n_\mu T^{\mu\nu} \beta_\nu - \xi n_\mu j^\mu.
\ee
The left hand side is the entropy density seen by the observer moving with four-velocity
$n$. If $n_\mu = \hat \beta_\mu$, the eq.~(\ref{smunmu}) is manifestly the basic 
relation of thermodynamics expressing the proper entropy density $s$ as a function 
of proper energy and charge density
\be\label{basic}
  \sqrt{\beta^2} s = \beta \cdot \phi + \beta_\mu \beta_\nu T^{\mu \nu} -
  \xi \beta_\mu j^\mu ,
\ee
provided that $\beta^2 = 1/T^2$ and $\xi= \mu/T$, what makes the physical meaning
of $\beta$ and $\xi$ apparent. Indeed, introducing the symbols $\rho$ for the proper 
energy density and $q$ for the proper charge density
\be\label{localtd}
   T s = T^2 \beta \cdot \phi + \rho - \mu q,
\ee
where 
\be\label{energydensity}
  \rho=\frac{\beta_\mu \beta_\nu T^{\mu\nu}}{\beta^2}\quad\quad 
   q=\frac{\beta_\mu j^\mu}{\sqrt{\beta^2}}.
\ee
The equation (\ref{localtd}) tells us that the $\beta$ frame is the one where the 
basic thermodynamic relation between proper entropy density and proper (true) energy 
and charge densities takes on its simplest form. In different frames, this relation 
is to be obtained contracting with a vector different from $n$ and it may thus 
contain additional terms, most likely of the second order in derivatives (see 
discussion in Appendix B). We conclude this section by noting that in the familiar 
global thermodynamical equilibrium, it is known \cite{israel,becacov} that the 
four-vector field $\phi^\mu = p\beta^\mu$ where $p$ is the pressure, hence the 
(\ref{localtd}) can be written in the more familiar form
\be\label{localtd2}
   T s = p + \rho - \mu n.
\ee
In fact, at local thermodynamical equilibrium, the thermodynamic potential current 
$\phi$ may have additional terms depending, e.g., on derivatives of the $\beta$ and 
$\xi$ fields. If these additional terms do have a longitudinal (along $\beta$) component, 
then the above equation is to be replaced by the most general (\ref{localtd}).

\section{Local thermodynamical equilibrium for a general $\beta$ field}
\label{local2}

For a general, non vorticity-free field $\beta$, the identification $n=\hat\beta$ 
is not possible and must be modified. One can iteratively construct a field $n(x)$
which fulfills eq.~(\ref{pfaff}) and, at the same time, reproducing the known features 
of global thermodynamical equilibrium with rotation (see discussion in \cite{becatinti1,
becacov}). Take:
\be\label{firstorder}
    b^{(1)}_\mu=\beta _\mu + \frac{1}{2} x^\nu \left( \partial_\mu \beta_\nu -  
    \partial_\nu \beta_\mu \right) 
\ee
Clearly, $\partial_\mu b^{(1)}_\nu-\partial_\nu b^{(1)}_\mu\approx O(\partial^2)$. 
Iteratively, one can add to $b^{(1)}_\mu$ higher order derivative terms which
are antisymmetric in $\mu\nu$ to eliminate gradients at some order. For instance,
\be\label{secondorder}
  b^{(2)}_\mu = \beta_\mu + x^\nu\left(1+\frac{x\cdot \partial}{3}\right)
  \left[ \frac{1}{2} \left( \partial_\mu \beta_\nu - \partial_\nu \beta_\mu \right)
  \right]  
\ee
implies $\partial_\mu b^{(2)}_\nu -\partial_\nu b^{(2)}_\mu\approx {\cal O}(\partial ^3)$. 
Thereby, we can construct a field $b(x)$ with vanishing external derivative
\be\label{bextdev}
 \partial_\mu b_\nu - \partial_\nu b_\mu = 0,
\ee
which can be used to define the LTE hypersurfaces $\Sigma(\tau)$, i.e:
$$
    n (x) \equiv \hat b(x).
$$
because, as it is apparent, if $b$ fulfills eq.~(\ref{pfaff}), any field collinear to 
it will. Hence, from eq.~(\ref{firstorder})-(\ref{secondorder}) we generalize to all 
orders, defining $\varpi$:
\be\label{kill}
   \beta_\nu(x) \equiv b_\nu (x) + \varpi_{\nu\lambda}(x) x^\lambda, 
\ee
with:
\be\label{thvort}
   \varpi_{\nu\lambda}(x) = -\frac{1}{2} (\partial_\nu \beta_\lambda - 
   \partial_\lambda \beta_\nu) - \frac{1}{6} \left( x^\rho \partial_\rho 
   \partial_\nu \beta_\lambda - x^\rho \partial_\rho \partial_\lambda \beta_\nu 
   \right) + \ldots
\ee
It is apparent that the obtained expressions (\ref{kill}) and (\ref{thvort}) are 
in full agreement with the global equilibrium case (\ref{glob}) and (\ref{glob2})
respectively. 

With the same LTE density operator as in eq.~(\ref{rhol2}), the field $\beta$ is 
now the solution of a modified version of the eq.~(\ref{betaeq2}) enforcing the equality 
of the mean-energy and momentum density
\bea\label{betaeq3}
  && b_\mu T^{\mu\nu}_{\rm LE}[\beta,\xi] =  b_\mu T^{\mu\nu}  \implies
  (\beta_\mu - \varpi_{\mu\lambda}[\beta] x^\lambda) T^{\mu\nu}_{\rm LE}[\beta,\xi]
  = (\beta_\mu - \varpi_{\mu\lambda}[\beta] x^\lambda) T^{\mu\nu} \nonumber \\
  && b_\mu j^{\mu}_{\rm LE}[\beta,\xi] =  b_\mu j^{\mu}  \implies
  (\beta_\mu - \varpi_{\mu\lambda}[\beta] x^\lambda) j^{\mu}_{\rm LE}[\beta,\xi]
  = (\beta_\mu - \varpi_{\mu\lambda}[\beta] x^\lambda) j^{\mu},
\eea
with $\varpi$ given by (\ref{thvort}). We stress that, for a vorticous $\beta$ field,
it is not possible to restore eq.~(\ref{betaeq2}) instead of (\ref{betaeq3}) to 
determine $\beta$, for $T_{\rm LE}$ requires a LTE density operator (\ref{rhol2}) 
to be defined and this in turn {\em demands} the constraints in the specific form 
(\ref{constr}) with $n$ vorticity-free.

By using (\ref{rhol2}) and (\ref{tpcurrent}), one can find an expression of 
entropy current
\be\label{current2}
   s^\mu = \phi^\mu + T^{\mu\nu}_{\rm LE} \beta_\nu - \xi j^\mu_{\rm LE} + 
   s_T^\mu(n).
\ee
The entropy density in the local rest frame of the fluid is obtained by contracting 
(\ref{current2}) with $\beta$
$$
 \sqrt{\beta^2} s = \phi^\mu \beta_\mu + \beta_\mu T^{\mu\nu}_{\rm LE} 
 \beta_\nu - \xi \beta_\mu j^\mu_{\rm LE} + \beta \cdot s_T(n).
$$
However, unlike in the non-vorticous case, replacing the local equilibrium mean 
values with the true ones is not straightforward.

\section{Temperature and thermometers in relativity}
\label{thermometer}

So far, we have defined the temperature (and four-velocity) in a local equilibrium 
state as a Lagrange multiplier in the constrained (with fixed energy and momentum 
densities) maximization of the entropy. 
This mathematical definition corresponds to a more physical one which can be obtained
by introducing the notion of an ideal relativistic thermometer. Just as in classical 
thermodynamics, this is, by definition, a ``small'' 
object able of instantaneaously achieve thermodynamical equilibrium with the system 
in contact with it. Besides, it should have some macroscopic internal property (such 
as size, resistivity etc.) which varies as a function of temperature, so that it can 
be used to define a scale thereof. 

In the relativistic context, an ideal thermometer can exchange both energy and momentum 
with the system, and therefore its response is not limited to a change of its internal
property gauging the temperature but it also includes a change of its four-velocity. 
In other words, once in contact with the system, the idealized relativistic thermometer 
will move at some finite speed which is determined by the local equilibrium conditions.
Now, the discussion gets easier considering both the system and the thermometer small yet 
finite. If the thermometer attains full thermodynamical equilibrium with the system, 
the entropy will be maximal with respect to energy and momentum exchange, thus we 
can write (the subscript ${\rm T}$ refers to the thermometer quantities)
\be\label{thermom1}
 \frac{\partial S}{\partial P^\mu} = \frac{\partial S_{\rm T}}{\partial P^\mu_{\rm T}},
\ee
keeping the proper volumes and the conserved charges fixed. Now, let us suppose that
the the system is so small that $\beta$ and $\xi$ are essentially constant over the 
system and thermometer volumes so as to taking them out of the integral sign in 
eq.~(\ref{entropy}) \footnote{Henceforth, we will use the shorthand $\di \Sigma_\mu$
for $\di \Sigma \, n_\mu$.}
$$
 S = \log Z_{\rm LE} + \int \di \Sigma_\mu \; \left( T^{\mu\nu} \beta_\nu 
 - \xi j^\mu \right) \simeq \log Z_{\rm LE} + \beta_\nu 
 \int \di \Sigma_\mu T^{\mu\nu} - \xi \int \di \Sigma_\mu j^\mu
 = \log Z_{\rm LE} + \beta_\nu P^\nu - \xi Q
$$
where we have used the (\ref{constr}). Note that $P$ and $Q$ do not depend on the 
frame because the divergences of $T$ and $j$ are assumed to vanish (interaction 
energy between system and thermometer is negligible by assumption). Hence, according 
to eq.~(\ref{thermom1}) and keeping in mind the basic relations of equilibrium 
relativistic thermodynamics which express the mean values of energy-momentum as
derivatives of $\log Z_{\rm LE}$ we obtain
$$
  \beta_\nu = \beta_{\nu {\rm T}}.
$$
The above equation implies that a relativistic thermometer in thermodynamical 
equilibrium with the system will mark the temperature $T_0=/\sqrt{\beta^2}$ and move 
with a speed $\beta/\sqrt{\beta^2}$. In this case, the thermometer is defined as 
{\em comoving} and the marked temperature is generally referred to as {\rm the} 
local temperature. 

Alternatively, one can retain a more traditional definition of an ideal thermometer 
as a ``small'' object endowed with a temperature gauge and able of instantaneaously 
achieve thermodynamical equilibrium with the system in contact with it {\em with 
respect to energy exchange}; its velocity $v$ can be externally imposed. According 
to the generally accepted extension of thermodynamics to relativity \cite{einplanck}, 
one has to choose the frame where the thermometer is at rest and therein enforce 
the condition of maximal entropy with respect to only {\em energy} exchange
\be
 \frac{\partial S}{\partial E} = \frac{\partial S_{\rm T}}{\partial E_{\rm T}},
\ee
which results in the equality of the time components of the $\beta$ vectors in that 
frame:
$$
  \beta^0 = \beta^0_{\rm T}
$$
or
$$
  \beta \cdot v = \frac{1}{T_{\rm T}}.
$$
In conclusion, a thermometer moving with four-velocity $v$ in a system in local 
thermodynamical equilibrium, characterized by a four-vector field $\beta$, will 
mark a temperature which is equal to
\be\label{temp}
 T_T = \frac{1}{\beta(x) \cdot v}.
\ee
As the scalar product of two timelike unit vectors $ u \cdot v \ge 1$ and
$$
    u \cdot v = 1 \qquad {\rm iff} \;\; u = v
$$    
one has, according to (\ref{temp})
$$
  T_T \le T_0 = \frac{1}{\sqrt{\beta^2}}  \qquad \qquad  T_T = T_0 \qquad {\rm iff} 
  \;\;  u = v,
$$
that is the temperature marked by an idealized thermometer is maximal if it moves with
the same four-velocity of the fluid. Thereby, we can establish a thought operational 
procedure to define a four-velocity of the fluid based on the notion of LTE at the 
spacetime point $x$:
\begin{itemize}  
\item{} put (infinitely many) ideal thermometers in contact with the relativistic 
system at the spacetime point $x$, each with a different four-velocity $v$;
\item{} the ideal thermometer marking the {\em highest} value $T_0$ moves, by 
definition, with the four-velocity $u(x)= T_0 \beta(x) = 1/\sqrt{\beta^2}$.
\end{itemize}

\section{The stress-energy tensor in the $\beta$ frame}
\label{setensor}

As has been mentioned in sect.~\ref{local}, the LTE density operator that we have 
defined and discussed in sect.~\ref{local} is not the true density operator $\wrho$. In 
the Heisenberg representation, the true density operator is stationary, time-independent, 
which is evidently not the case for $\wrhol$ in eq.~(\ref{rhol2}), which depends 
on time $\tau$ so as to the total entropy can change (in fact increase) in time. 
The true stationary density operator $\wrho$ is the one needed to write the continuity 
equations of the mean values of operators, such as the stress-energy tensor:
\be\label{trueset}
  \partial_\mu T^{\mu\nu} = \partial_\mu \tr (\wrho \, \wT^{\mu\nu})_{\rm ren} = 
  \tr (\wrho \, \partial_\mu \wT^{\mu\nu})_{\rm ren} = 0.
\ee
The  eq.~(\ref{trueset}) is the basic equation of relativistic hydrodynamics and, in
the above form, makes it clear that the conservation of the mean value stems from 
the more fundamental conservation equation of the corresponding quantum operator.

If, at some initial time $\tau_0$, the system is known to be at local thermodynamical 
equilibrium, one can take the actual, time-independent, density operator as the 
one in eq.~(\ref{rhol2}) provided that both the spacelike hypersurface $\Sigma$ and the 
operators $\wT$, $\wj$ are evaluated at $\tau_0$:
\be\label{tevol}
 \wrho = \dfrac{1}{Z} 
 \exp\left[ - \int_{\Sigma(\tau_0)} \!\!\!\!\!\! \di \Sigma \; n_\mu \left( \wT^{\mu\nu} 
  \beta_\nu - \wj^\mu \xi \right)\right].
\ee
Consider now the evolution in $\tau$ of the LTE hypersurface $\Sigma$; one can then
rewrite the density operator in (\ref{tevol}) in terms of the operators at the 
present time $\tau$ by means of the Gauss' theorem:
\be\label{tevol2}
 \wrho = \dfrac{1}{Z} 
 \exp\left[ - \int_{\Sigma(\tau)} \!\!\!\!\!\! \di \Sigma \; n_\mu \left( \wT^{\mu\nu} 
  \beta_\nu - \wj^\mu \xi \right) + \int_\Omega \di \Omega \; \left( \wT^{\mu\nu} d_\mu 
  \beta_\nu - \wj^\mu d_\mu \xi \right) \right], 
\ee
where $d$ stands for the covariant derivative in the coordinates $\tau$ and $\sigma_i \;
i=1,2,3$ of the surfaces $\Sigma$. The region $\Omega$ is the portion of spacetime 
enclosed by the two hypersurface $\Sigma(\tau_0)$ and $\Sigma(\tau)$ and the timelike 
hypersurface at their boundaries, where the flux of ($\wT^{\mu\nu} \beta_\nu(x) - 
\wj^\mu \xi(x)$) is supposed to vanish (see e.g. fig.~\ref{sigmas}). 

The first term of the exponent on the right hand side of the eq.~(\ref{tevol2}) is 
just the LTE exponent at time $\tau$. If the evolution of the stress energy tensor 
and current operators are such that the system keeps close to a situation of local 
thermodynamical equilibrium - a request of relativistic hydrodynamics - the second 
term in the exponent 
can be considered as a perturbation with respect to the first term and, accordingly, 
an expansion can be made in the gradients of the $\beta$ and $\xi$ fields with the 
method of linear response theory, through an iterated use of the operator Kubo identity.
This in essence, is the method put forward by Zubarev \cite{zuba} and used by A.~Hosoya 
{\em et al.} \cite{hosoya} to generate so-called Green-Kubo formulae of transport
coefficients for a relativistic fluid, which coincide with those obtained by using
the method of the variation of the metric into the equilibrium euclidean action 
\cite{Baier:2007ix}. The expansion allows to express the mean value of a spacetime-dependent 
operator $\widehat O(x)$ with $x=(\tau,\sigma)$ as the mean value at LTE plus a 
correction depending on the gradients:
\be\label{kuboexp}
  \langle \widehat O (x) \rangle \simeq \langle \widehat O (x) \rangle_{\rm LE} 
  - \langle \widehat O (x) \rangle_{\rm LE} \langle \widehat B \rangle_{\rm LE}
  + \int_0^1 \di z \; \langle \widehat O (x') \e^{z\widehat A} \widehat B 
  \e^{-z\widehat A} \rangle_{\rm LE},
\ee
choosing the LTE hypersurface going through the point $x$. In eq.~(\ref{kuboexp})
the operators $\widehat A$ and $\widehat B$ are, respectively, the first and the second 
integral in the exponent of eq.~(\ref{tevol2}). In flat spacetime, the integration 
region $\Omega$ is bounded by the two LTE hypersurfaces at $\tau$ and $\tau_0$. 
They can be approximated by the spacelike tangent hyperplanes at the points 
$x=(\tau,\sigma)$ and $(\tau_0,\sigma)$ respectively, whose normal versor is 
$n_\mu$. This allows to carry out the integration over Minkowski spacetime, with 
the time $t$ marked by an observer moving with velocity $n$, as well as replacing 
covariant derivative with usual derivatives:
$$
  \int_\Omega \di \Omega \; \left( \wT^{\mu\nu} d_\mu \beta_\nu - \wj^\mu 
  d_\mu \xi \right) \rightarrow \int_{T\Omega} \di^4 x \; \left( \wT^{\mu\nu} 
  \partial_\mu \beta_\nu - \wj^\mu \partial_\mu \xi \right) .
$$
Altogether, this approach generates an expansion of the stress-energy tensor (as
well as any operator) from the LTE point in the gradients of the thermodynamic fields
$\beta$ and $\xi$ which is - as we will see - equivalent to that in the usual $u$,
$T$ and $\mu$:
\be\label{setexpa}
  T^{\mu\nu} = \tr (\wrho \wT^{\mu\nu})_{\rm ren} = T^{\mu\nu}_{\rm LE}(x)
  + \delta T^{\mu\nu}(\partial\beta,\partial\xi).
\ee

However, neither the hydrodynamical frame nor the zero-order term of the expansion, 
that is the mean value at LTE, were discussed in detail in ref.~\cite{hosoya}, 
where it was simply assumed that $T^{\mu\nu}_{\rm LE}(x)$ has the familiar ideal
form:
\be\label{tideal}
  T_{\rm id}^{\mu\nu}(x) = (\rho + p)_{\rm eq} \frac{1}{\beta^2} \beta^\mu(x) 
  \beta^\nu(x) - p_{\rm eq} g^{\mu\nu}.
\ee
In fact, as we will see, the zero-order term, that is:
\be\label{leqse}
  T^{\mu\nu}_{\rm LE}(x) = \tr ( \wrhol \wT^{\mu\nu}(x))_{\rm ren} = 
  \frac{1}{Z_{\rm LE}} \tr \left( \exp \left[ -\int \di \Sigma_\mu \; \left( 
  \wT^{\mu \nu}\beta_\nu - \xi \wj^\mu \right) \right] \wT^{\mu\nu}(x) \right)_{\rm ren},
\ee  
is less trivial than generally believed and the choice of a hydrodynamical frame is 
crucial to determine its value. This is the subject of the remaining part of this
section

\subsection{The stress-energy tensor at local thermodynamical equilibrium}

We first remark that, being $\beta$ a function of the spacetime point, the trace in 
(\ref{leqse}) cannot be calculated straightforwardly. However, in the exponent 
of $\wrhol$, one can make a Taylor expansion in $\beta$ and $\xi$ about the same 
point $x$ where the stress-energy tensor is to evaluated. The idea is that, at LTE, 
only the nearby points will contribute to its mean value, especially if the gradients 
are small. In other words, in the so-called hydrodynamical limit, the $\beta$ field 
is mostly uniform in the region where the stress-energy tensor correlation function,
determined by microscopic correlation lengths, is significant. Hence:
\bea\label{leqexpa}
  && \exp \left[ -\int \di \Sigma_\mu \; \left( \wT^{\mu \nu}\beta_\nu - 
  \xi \wj^\mu \right) \right] \nonumber \\
 \simeq && \exp \left[ - \beta_\nu(x) \int \di \Sigma_\mu \; \wT^{\mu \nu} + 
 \xi(x) \int \di \Sigma_\mu \; \wj^\mu - \frac{\partial \beta_\nu}{\partial \sigma_i}(x) 
  \int \di \Sigma_\mu \; \wT^{\mu \nu} (\sigma_i - \sigma_{0i}) + 
  \frac{\partial \xi}{\partial \sigma_i}(x) \int \di \Sigma_\mu \; \wj^\mu 
  (\sigma_i - \sigma_{0i}) + \ldots \right] \nonumber \\
 = && \exp \left[- \beta_\nu(x) \widehat P^\nu + \xi(x) \widehat Q 
 - \frac{\partial \beta_\nu}{\partial \sigma_i}(x) 
  \int \di \Sigma_\mu \; \wT^{\mu \nu} (\sigma_i - \sigma_{0i}) + 
  \frac{\partial \xi}{\partial \sigma_i}(x) \int \di \Sigma_\mu \; \wj^\mu 
  (\sigma_i - \sigma_{0i}) + \ldots \right],
\eea
where $\sigma$ are the curvilinear coordinates of the hypersurface $\Sigma$ at the 
time $\tau$ (the point $x$ has coordinates $\tau$ and $\sigma_0$). In the last equality 
we have taken into account that the integrals of the stress-energy tensor and the 
current over any 3D hypersurface equal the total four-momentum and charge. Now
$$
 \sum_{i=1}^3 \frac{\partial \beta_\nu}{\partial \sigma_i}(x) (\sigma_i - \sigma_{0i})
 = \sum_{i=1}^3 \partial_\lambda \beta_\nu (x) \frac{\partial x^\lambda}{\partial \sigma_i}(x) 
 (\sigma_i - \sigma_{0i}) = \partial_\lambda \beta_\nu (x) 
 \sum_{i=1}^3  t^\lambda_i(x) (\sigma_i - \sigma_{0i}),
$$
where $t_\mu^i$ are the vectors tangent to the hypersurface $\Sigma$. If the $\beta$
field is vorticity-free, one can choose the $\beta$ frame with $n = \hat \beta$,
thus the vectors $t^i$ will be simply orthogonal to $\beta$. Hence, denoting with
$y$ the point with coodinates $\tau$ and $\sigma$
$$
\sum_{i=1}^3  t^\lambda_i(x) (\sigma_i - \sigma_{0i}) \simeq (y_\lambda - x_\lambda)_T,
$$
where the subscript $T$ stands for the transverse projection with respect to $\beta$; 
introducing the definitions
\be\label{derdef}
 D \equiv u^\mu \partial_\mu = \frac{1}{\sqrt{\beta^2}} \beta^\mu \partial_\mu
 = T \beta^\mu \partial_\mu
 \qquad \qquad \nabla^\nu \equiv (g^{\mu\nu} - u^\mu u^\nu) \partial_\mu =
 (g^{\mu\nu} - T^2 \beta^\mu \beta^\nu) \partial_\mu \equiv \Delta^{\mu\nu} 
 \partial_\mu
\ee
where $T = 1/\sqrt{\beta^2}$ is the comoving temperature, one can finally rewrite 
the eq.~(\ref{leqexpa}) as
\bea\label{leqexpa2}
 && \exp \left[ -\int \di \Sigma_\mu \; \left( \wT^{\mu \nu}\beta_\nu - 
  \xi \wj^\mu \right) \right] \nonumber \\  
 \simeq && \exp \left[- \beta_\nu(x) \widehat P^\nu + \xi(x) \widehat Q 
  - \partial_\lambda \beta_\nu(x) \int_{T\Sigma} \di \Sigma_\mu(y) \; \wT^{\mu \nu}(y) 
  (y^\lambda - x^\lambda)_T + \partial_\lambda \xi(x) \int_{T\Sigma} \di \Sigma_\mu(y) \; 
  \wj^\mu(y) (y^\lambda - x^\lambda)_T + \ldots \right] \nonumber \\
 = && \exp \left[- \beta_\nu(x) \widehat P^\nu + \xi(x) \widehat Q 
  - \nabla_\lambda \beta_\nu(x) \int_{T\Sigma} \di \Sigma_\mu(y) \; \wT^{\mu \nu}(y) 
  (y^\lambda - x^\lambda)_T + \nabla_\lambda \xi(x) \int_{T\Sigma} \di \Sigma_\mu(y) \; 
  \wj^\mu(y) (y^\lambda - x^\lambda)_T + \ldots \right] \nonumber \\
 = && \exp \left[- \beta_\nu(x) \widehat P^\nu + \xi(x) \widehat Q 
  - \frac{1}{4} (\nabla_\lambda \beta_\nu(x) - \nabla_\nu \beta_\lambda(x)) 
  \widehat J^{\lambda\nu}_{xT} + \frac{1}{2} (\nabla_\lambda \beta_\nu(x) + 
  \nabla_\nu \beta_\lambda(x)) \widehat L^{\lambda\nu}_x + \nabla_\lambda \xi(x) 
  \widehat d^\lambda_x + \ldots \right],
\eea
where the integration - to a good approximation - can be carried out on the hyperplane 
$T\Sigma$ tangent to $\Sigma$ at the point $x$. In the eq.~(\ref{leqexpa2}), the 
operator $\widehat J_{xT}$ is the transverse projection of the angular momentum 
operator around the point $x$:
$$
  \widehat J^{\lambda\nu}_{xT} \equiv \int_{T\Sigma} \di \Sigma(y) \; n_\mu 
  (y^\lambda-x^\lambda)_T \, \wT^{\mu\nu}(y) - (\lambda \leftrightarrow \nu) 
$$   
and:
\bea\label{opdef}
  && \widehat L^{\lambda\nu}_x \equiv \frac{1}{2} \int_{T\Sigma} \di \Sigma(y) 
   \; n_\mu (y^\lambda-x^\lambda)_T \,
  \wT^{\mu\nu}(y) +(\lambda \leftrightarrow \nu) \nonumber \\
  && \widehat d^{\lambda}_x \equiv \int_{T\Sigma} \di \Sigma(y) \; 
  n_\mu (y^\lambda-x^\lambda)_T \, \wj^{\mu}(y).
\eea
We note in passing that $\widehat L_x$ and $\widehat d_x$ are {\em not} true tensors 
because they are integrals of non-conserved densities; their definition is only
valid for the specific frame. 

Unfortunately, the expression (\ref{leqexpa2}) does not imply the full correct global 
equilibrium limit (\ref{globrho}). Particularly, it can be realized that this happens
only if, at the global equilibrium defined by the equations (\ref{glob}),(\ref{glob2})
one has:
$$
  \varpi_{\lambda\nu} \beta^\lambda = 0
$$
However, in the global equilibrium with rotation (see sect.~\ref{frames}), $\varpi_{\mu\nu} 
\beta^\nu \ne 0$ and proportional to the acceleration field. The reason of this 
shortcoming is the choice of $n = \hat \beta$, which is possible, as has been mentioned, 
only if the $\beta$ field is vorticity-free, what is not true even for the simple 
case of rotating global equilibrium. 

To find the correct expression it is convenient to use the decomposition in eq.~(\ref{kill}) 
to rewrite the integral in the exponent of LTE density operator (\ref{rhol2}) as:
\begin{eqnarray*}
 -\int \di \Sigma \; n_\mu \wT^{\mu\nu} \beta_\nu && =
 -\int \di \Sigma(y) \; n_\mu \wT^{\mu\nu} (b_\nu + \varpi_{\nu \lambda} y^\lambda) 
 \nonumber \\
 && =  -\int \di \Sigma(y) \; n_\mu \left[ \wT^{\mu\nu} b_\nu - \frac{1}{2} 
 \varpi_{\lambda\nu} (y^\lambda  \wT^{\mu\nu} - y^\nu  \wT^{\mu\lambda}) \right]
\end{eqnarray*}
where we have set $\xi=0$ for simplicity. We can now make a first-order Taylor expansion 
of the thermodynamic field $b$ in the integrand about the point $x$ and replace the
integration domain with the hyerplane $T\Sigma$ tangent to $\Sigma$ in $x$ if necessary:
\bea\label{expo}
 && -\int \di \Sigma(y) \; n_\mu \left[ \wT^{\mu\nu} b_\nu -\frac{1}{2} \varpi_{\lambda\nu} 
 (y^\lambda  \wT^{\mu\nu} - y^\nu  \wT^{\mu\lambda}) \right] \nonumber \\
 \simeq && - b_\nu(x) \int \di \Sigma(y) \; n_\mu \wT^{\mu\nu} - \frac{\partial b_\nu}{\partial x^\rho}
 \int_{T\Sigma} \di \Sigma(y) \; n_\mu (y^\rho-x^\rho)_T \wT^{\mu\nu} + \frac{1}{2} \int \di \Sigma(y) 
 \; n_\mu (y^\lambda  \wT^{\mu\nu} - y^\nu  \wT^{\mu\lambda}) \varpi_{\lambda\nu}(y)
 \eea
where the subscript $T$ now stands for orthogonal to the vector $b(x)$. The integral in
the first term on the right hand side of the above equation is just the four-momentum
$\widehat P^\nu$, while the second term can be decomposed into symmetric and antisymmetric 
contributions. Since the $b$ field has a vanishing antisymmetric gradient 
(see eq.~(\ref{bextdev}), one is left with:  
\be\label{expo2}
  - b_\nu(x) \widehat P^\nu - \frac{1}{4} (\partial_\rho b_\nu + \partial_\nu b_\rho )
 \int_{T\Sigma} \di \Sigma(y) \; n_\mu \left[ (y^\rho-x^\rho)_T \wT^{\mu\nu} + (y^\nu-x^\nu)_T 
 \wT^{\mu\rho} \right] + \frac{1}{2} \int \di \Sigma(y) 
 \; n_\mu (y^\lambda  \wT^{\mu\nu} - y^\nu  \wT^{\mu\lambda}) \varpi_{\lambda\nu}(y)
\ee
where we have used the orthogonality between the tangent vectors to $\Sigma$ and $b$
implied by the choice $n= \hat\beta$ and the fact that $b$ field has vanishing 
external derivative (see eq.~(\ref{bextdev})), thus only the symmetric combination
of integral and derivatives of $b$ in eq.~(\ref{expo}) is retained.
 
We now want to work out and further expand the (\ref{expo2}) so as to have in it
only linear terms in the first order $\beta$ derivatives. As a first step, we can 
Taylor expand the tensor $\varpi$ in the last integral expression in eq.~(\ref{expo2}) 
about the same point $x$ as for $b$; because of the (\ref{thvort}), in this 
expansion we will only retain the zeroth order term if second order derivatives of 
$\beta$ are not to appear and approximate $\varpi(x)$ with the antisymmetric part of the 
$\beta$ gradient in $x$. Secondly, we note that, according to the definition (\ref{kill}) 
and the eq.~(\ref{thvort}), the symmetric part of the gradient of $b$ differs from 
the corresponding symmetric part of the gradient of $\beta$ by terms involving 
higher order derivatives:
$$
 \partial_\rho b_\nu + \partial_\nu b_\rho = \partial_\rho \beta_\nu + 
 \partial_\nu \beta_\rho + {\cal O}(\partial^2).
$$
Therefore, we can rewrite eq.~(\ref{expo2}) as:
\begin{eqnarray*}
 && - b_\nu(x) \widehat P^\nu - \frac{1}{4} (\partial_\rho \beta_\nu + \partial_\nu \beta_\rho )
 \int_{T\Sigma} \di \Sigma(y) \; n_\mu \left[ (y^\rho-x^\rho)_T \wT^{\mu\nu} + (y^\nu-x^\nu)_T 
 \wT^{\mu\rho} \right] + \frac{1}{2} \varpi_{\lambda\nu}(x) \int \di \Sigma(y) 
 \; n_\mu (y^\lambda  \wT^{\mu\nu} - y^\nu  \wT^{\mu\lambda}) \nonumber \\
 && =  - b_\nu(x) \widehat P^\nu - \frac{1}{4} (\partial_\rho \beta_\nu + \partial_\nu \beta_\rho )
 \int_{T\Sigma} \di \Sigma(y) \; n_\mu \left[ (y^\rho-x^\rho)_T \wT^{\mu\nu} + (y^\nu-x^\nu)_T 
 \wT^{\mu\rho} \right] + \frac{1}{2} \varpi_{\lambda\nu}(x) \widehat J^{\lambda\nu}
\end{eqnarray*}
The integrand argument $(y^\rho-x^\rho)_T$ is transverse to $b$ in the point $x$, 
which means that it is also transverse to $\beta$ up to first order derivatives of 
$\beta$ in view of eq.~(\ref{glob}). Hence:
$$
 \Delta_b^{\mu\nu}(x) = g^{\mu\nu} - \frac{b^\mu(x) b^\nu(x)}{b^2(x)} 
 \simeq g^{\mu\nu} - \frac{\beta^\mu(x) \beta^\nu(x)}{\beta^2(x)} + 
 {\cal O}(\partial \beta) \simeq \Delta_\beta^{\mu\nu}(x)
$$
We can then replace the transverse projector on the hypersurface orthogonal to $b$
with the one transverse to $\beta$ in eq.~(\ref{opdef}) and write:
\be\label{expo3}
  - b_\nu(x) \widehat P^\nu + \frac{1}{2} \varpi_{\lambda\nu}(x) \widehat J^{\lambda\nu}
  - \frac{1}{2} (\partial_\lambda \beta_\nu + \partial_\nu \beta_\lambda) 
  \widehat L^{\lambda\nu}_x  
\ee  
where $\widehat L_x$ is defined in eq.~(\ref{opdef}). Finally, by using the identity
(\ref{angmomshift}) and the relation (\ref{kill}), one can rewrite eq.~(\ref{expo3}) as:
\be\label{expo4}
 - \beta_\nu(x) \widehat P^\nu + \frac{1}{2} \varpi_{\lambda\nu}(x) \widehat J_x^{\lambda\nu}
 - \frac{1}{2} (\partial_\lambda \beta_\nu + \partial_\nu \beta_\lambda) \widehat L^{\lambda\nu}_x 
\ee  
and, finally, restoring the chemical potential term and replacing $\varpi$ with
its first order approximation in the $\beta$ derivatives: 
\be\label{corrected}
 \wrhol \simeq \frac{1}{Z_{\rm LE}}\exp \left[- \beta_\nu(x) \widehat P^\nu + 
 \xi(x) \widehat Q  - \frac{1}{4} (\partial_\nu \beta_\lambda(x) - \partial_\lambda 
 \beta_\nu(x)) \widehat J_x^{\lambda\nu} + \frac{1}{2}(\partial_\nu \beta_\lambda(x) 
 + \partial_\lambda \beta_\nu(x)) \,\widehat L^{\lambda\nu}_x + \nabla_\lambda \xi(x) 
 \, \widehat d^\lambda_x \right].
\ee
It can be seen that this expression has the correct global equilibrium limit in 
eq.~(\ref{glob}): as has been mentioned, the coefficient of $\widehat L$ and $\widehat d$
vanish because of the eq.~(\ref{killing}) and $\varpi = const$ is given by the external 
derivative of the $\beta$ field like in the eq.~(\ref{thvort}). 

The expression (\ref{corrected}), once (\ref{thvort}) is taken into account, implies 
that $\wrhol$ can again be expanded in the gradients of the $\beta$ and $\xi$ fields 
with linear response theory starting from a point of global thermodynamical 
equilibrium with constant inverse four-temperature $\beta(x)$ and chemical potential 
$\xi(x)T(x)$, where
$$
     \wrho_{\rm eq} = \frac{1}{Z_{\rm eq}} \tr \left(
     \exp \left[- \beta_\nu(x) \widehat P^\nu + \xi(x) \widehat Q \right]\right). 
$$
Therefore
$$
  T^{\mu\nu}_{\rm LE}(x) \simeq \frac{1}{Z_{\rm eq}(\beta(x),\xi(x))}
  \tr \left( \exp \left[-\beta_\nu(x) \widehat P^\nu  + \xi(x) \widehat Q \right]
  \wT^{\mu\nu}(x) \right)_{\rm ren} + {\cal O}(\partial \beta, \partial \xi).
$$ 
The first term of the expansion can be readily identified: it is the mean value 
of the stress-energy tensor at the global thermodynamical equilibrium with a 
{\em global} inverse temperature four-vector and chemical potential equal to those
in $x$. In other words, it is the ideal part of the stress-energy tensor and the 
above expansion can be written as
\be\label{ideal}
  T^{\mu\nu}_{\rm LE}(x) \simeq T_{\rm id}^{\mu\nu}(x) + {\cal O}((\partial \beta, 
  \partial \xi)^N)
  = (\rho + p)_{\rm eq} \frac{1}{\beta^2} \beta^\mu(x) \beta^\nu(x) - p_{\rm eq} 
  g^{\mu\nu} + {\cal O}((\partial \beta, \partial \xi)^N),
\ee 
where the energy density $\rho$ and pressure $p$ are the same thermodynamic functions 
of $\beta^2(x),\xi(x)$ as at equilibrium. The eq.~(\ref{ideal}) shows that the 
mean value of the stress-energy tensor differs from the ideal one by terms which,
potentially, are of the first order in the gradients of $\beta$ and $\xi$. 

We believe, though we do not present here any calculation, that first-order terms 
in the gradient expansion of the mean value at LTE are vanishing, owing to general 
symmetry requirements. Instead, second order terms in the expansion of the operator 
(\ref{corrected}) should be non vanishing, whence $N=2$ in (\ref{ideal}). Some of
the coefficients in the second-order gradient expansion have been recently calculated 
in ref.~\cite{moore}.
Therefore, our terms obtained from an expansion of the LTE expression would either 
coincide with them - specifically the non-dissipative which survive, e.g., in the 
global equilibrium rotating case proportional to $\varpi \varpi$ - or additionally 
contribute to the second order dissipative coefficients, specifically those proportional 
to $\sigma \sigma$ or $\sigma\varpi$ where $\sigma$ is the symmetric part of 
$\partial_\mu\beta_\nu$. This will be the subject of further work.

\section{The $\beta$ frame vs Landau frame}
\label{frames}

We now come to a major point, namely the discussion of the difference between the 
$\beta$ frame and the familiar Landau frame. In the previous section we have seen
that, in the $\beta$ frame, eq.~(\ref{setexpa}) holds, and an equivalent one 
also holds for $j$:
$$
   T^{\mu\nu} = T^{\mu\nu}_{\rm LE} + \delta T^{\mu\nu} \qquad\qquad 
   j^{\mu} = j^{\mu}_{\rm LE} + \delta j^{\mu}.
$$
If the $\beta$ field is non-vorticous, then, because of the eq.~(\ref{betaeq2})
\be\label{conditions}
  \beta_\mu \delta T^{\mu\nu} = 0\quad\quad \beta_\mu \delta j^\mu = 0.
\ee
Indeed, the first of the two equations (\ref{conditions}) apparently imposes the 
orthogonality between the viscous part of the stress-energy tensor and the velocity 
vector, a condition often referred to as ``Landau matching condition'', so naively 
one would say that the $\beta$ frame and the Landau frame are equivalent, at least
as long as $\beta$ is vorticity-free. However, the actual definition of the Landau frame 
prescribes that the velocity four-vector $u_L$ is the timelike eigenvector of $T$
\be\label{landau}
   T^{\mu \nu} u_{L\nu} = \lambda u^\mu_L.
\ee       
It is worth remarking that the above Landau frame definition provides 4 independent 
equations, whereas the definition of the $\beta$ frame involve 5 equations. In fact, 
the Landau frame definition is usually, often tacitly, supplemented by the equality 
of the proper charge density respectively with its local equilibrium value
$$
   u_L \cdot j = u_L \cdot j_{\rm LE},
$$
which indeed amounts to enforce the second equality in the eq.~(\ref{conditions}).
In the traditional Landau scheme, this equation is sometimes justified through a 
redefinition of the temperature and chemical potential \cite{kovtun} in a non-equilibrium 
situation. However, as we have emphasized in this work, temperature and chemical 
potential can be unambiguously defined at the LTE, see sect.~\ref{local}. In fact, 
when changing frames, it should always be checked whether the basic relations involving 
thermodynamical quantities hold with the accordingly defined temperature and chemical 
potential.

The equation (\ref{conditions}) implies the eq.~(\ref{landau}) only if $\beta$ is 
an eigenvector of $T^{\mu\nu}_{\rm LE}$
$$
 0 = \beta_\mu \delta T^{\mu\nu}  = \beta_\mu (T^{\mu \nu} - T^{\mu \nu}_{\rm LE})
 = \beta_\mu T^{\mu \nu} - \lambda \beta^\nu,
$$
whence $\beta$ is the timelike eigenvector of $T$, so $\hat \beta \equiv u_L$.\\
So, {\em the $\beta$ frame coincides with the Landau frame if $\beta$ is vorticity
free and if it is the timelike eigenvector of $T_{\rm LE}$}. In all other cases, 
including the case of a vorticous $\beta$ field, the Landau and $\beta$ frame 
are not equivalent.
\begin{center}
\begin{figure}[ht]
\includegraphics[width=0.45\textwidth]{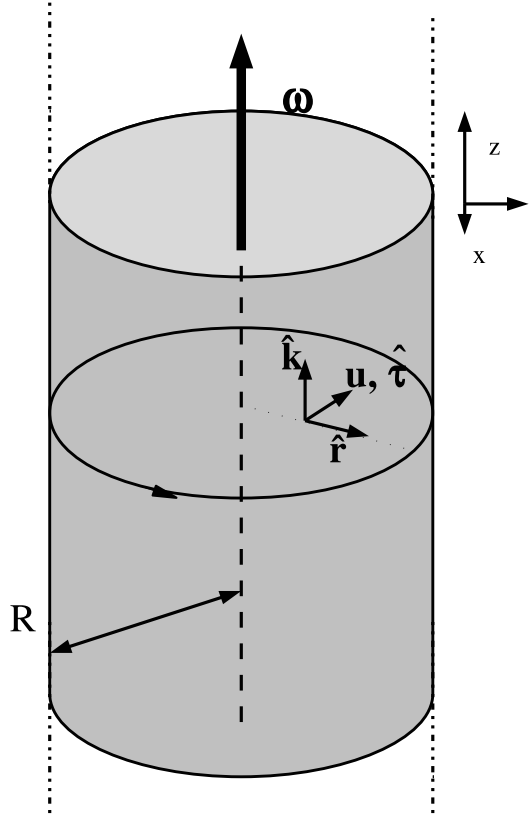}
\caption{Rotating cylinder with finite radius $R$ at temperature $T$. Also shown
the inertial frame axes and the spatial parts of the vectors of tetrad.
\label{illu}}
\end{figure}
\end{center} 

It can be readily realized that $\beta$ is an eigenvector of $T_{\rm LE}$ if 
$T_{\rm LE} = T_{\rm id}$. However, we have seen at the end of sect.~\ref{setensor} 
that this is not generally the case for the quantum form of LTE, i.e. there may be 
corrections to the ideal stress-energy tensor depending on the gradients of the 
$\beta$ field itself whose leading terms are expected to be quadratic. 

We are now going to discuss in detail a remarkable instance of of inequivalence between 
Landau and $\beta$ frames: the rotational ensemble, which is a global equilibrium case. 
Its density operator can be obtained from the eq.~(\ref{globrho}) by setting:
\be\label{thermpar} 
  b = (1/T_0,0,0,0) \qquad \qquad \varpi_{\mu\nu} = (\omega/T_0) (g_{1\mu} g_{2\nu} 
  - g_{1\nu} g_{2\mu})
\ee
that is:
\be\label{rotrho}
 \wrho = \frac{1}{Z} \exp[-\widehat H/T_0 + \omega \widehat J_z/T_0] \Pro_V,
\ee
where $\widehat J_z$ is the angular momentum operator along some fixed axis $z$
and $\omega$ has the physical meaning of a constant angular velocity 
(see fig.~\ref{illu}); $\Pro_V$ is a projector onto localized states, those obtained 
by enforcing peculiar boundary conditions on the quantum fields at some radius $R$ 
of an indefinitely long cylinder with axis $z$ and such that $\omega R < c$ (see
~\cite{becatinti1}). With the above choice of $b$ and $\varpi$, the relevant $\beta$ 
field in eq.~(\ref{glob}) reads:
$$
  \beta = \frac{1}{T_0} (1, \omegav \times {\bf x})
$$
where $\omegav = \omega \hat{\bf k}$. Its field lines are then circles centered on 
the $z$ axis (see fig.~\ref{illu}). Note that $1/\sqrt{\beta^2}\equiv T \ne T_0$, that
is the proper temperature differs from the constant "global" temperature $T_0$, a
well known relativistic feature. The density operator (\ref{rotrho}) is independent 
of the spacelike hypersurface $\Sigma$ (i.e. time-independent) provided that the flux 
of $\wT^{\mu\nu} \beta_\nu$ and $\wj^\mu$ vanish at the boundary
\be\label{rhoflux}
  \int_{\rm Boundary} \di \Sigma_\mu \left( \wT^{\mu\nu} \beta_\nu - 
  \xi \wj^\mu \right) = 0.
\ee  

In the usual formulation of relativistic hydrodynamics in the Landau frame the 
stress-energy tensor is decomposed along the $u_L$ vector as follows:
\be\label{landaudec}
  T^{\mu\nu} = (\rho_L + p) u^\mu_L u^\nu_L - p g^{\mu\nu} + \Pi^{\mu \nu},
\ee
with $\Pi^{\mu\nu} u_{L\nu} = 0$ by definition of $u_L$ but with the understood 
assumption that $\Pi \to 0$ at LTE \cite{MurongaPai1} and, {\it a fortiori}, at 
the global thermodynamical equilibrium. We will show that in the rotational case 
one has $\hat\beta \ne u_L$ as well as $\Pi \ne 0$. 

The latter inequality is expected to be a consequence of the fact that the density 
operator (\ref{rotrho}) has a cylindrical symmetry along $z$ axis, but not a full 
rotational symmetry, so there is no principle reason why the mean stress-energy 
tensor ought to be isotropic in its local (Landau) rest frame, or, in other words, 
why its spacelike eigenvalues ought to be the same. Indeed, with a cylindrical 
symmetry its most general form reads:
\bea\label{setrot}
  T^{\mu \nu} =  && G(r) u^\mu u^\nu + H(r) (\hat\tau^\mu u^\nu + \hat\tau^\nu u^\mu) +
  I(r) (\hat r^\mu u^\nu +\hat r^\nu u^\mu) \nonumber \\
  && + J(r) \hat\tau^\mu \hat\tau^\nu + K(r) (\hat r^\mu 
  \hat\tau^\nu +\hat r^\nu \hat\tau^\mu) + L(r)\hat r^\mu \hat r^\nu - M(r) g^{\mu \nu}, 
\eea
where $G, H, I, J, K, L, M$ are generic function of the radial coordinate $r$ such that  
$H(0)=I(0)=K(0)=0$, $u = \hat\beta$, $\hat r = (0,{\bf \hat r})$, $\hat k=(0,{\bf 
\hat k})$ and $\hat\tau$ is the spacelike versor orthogonal to the previous three, that is
\be\label{tauvers}
  \hat\tau = (\gamma v,\gamma {\bf \hat v})
\ee  
being ${\bf v} = \omegav \times {\bf x}$ and $\gamma = (1 - v^2)^{-1/2}$ (see 
fig.~\ref{illu}). The condition $\partial_\mu T^{\mu\nu} = 0$ allows to eliminate
either $I(r)$ or $K(r)$ and entails some differential relations between the functions 
in eq.~(\ref{setrot}). Clearly, if in the eq.~(\ref{setrot}) either $H$ or $I$ or both
are non-vanishing, the four-vector $u=\hat \beta$ is not an eigenvector of $T$ and the 
Landau and $\beta$ frame differ. Furthermore, if the scalar functions in (\ref{setrot}) 
do not meet specific relations, the diagonal form of the tensor at equilibrium is not 
the ideal one (no isotropy) and the understood assumption $\Pi = 0$ at equilibrium 
for the decomposition (\ref{landaudec}) breaks down. 
The rotational ensemble gives the opportunity to discuss in more detail the relevance
of these effects, namely the magnitude of the difference between $\hat\beta$ and $u_L$ 
and the relevant scales. As we pointed out at the end of sec.~\ref{setensor}, the 
leading global equilibrium corrections to the stress energy tensor are quadratic in 
the tensor $\varpi$, that is the antisymmetric part of $\partial \beta$. In natural 
units this tensor is adimensional and its magnitude in the rotational case, as implied 
by the eq.~(\ref{thermpar}) is $\hbar\omega/KT_0$ (natural constants are purposely restored
here). This means that the deviation from the familiar hydrodynamical scheme scales like
$(\hbar\omega/KT_0)^2$ which is a tiny number in most cases, still it can become relevant 
in special circumstances. It is worth stressing that $\hbar\omega/KT_0$, tiny as it might
be, is a further macroscopic scale independent of the microscopic scales such as a 
correlation length or a thermal wavelength. We can otherwise say that the stress-energy 
tensor deviations from the ideal form at equilibrium are to be expected in presence of 
a {\em local acceleration} such as when the velocity field is rigid. This will be the 
subject of further work.
\begin{center}
\begin{figure}[ht]
\includegraphics[width=0.7\textwidth]{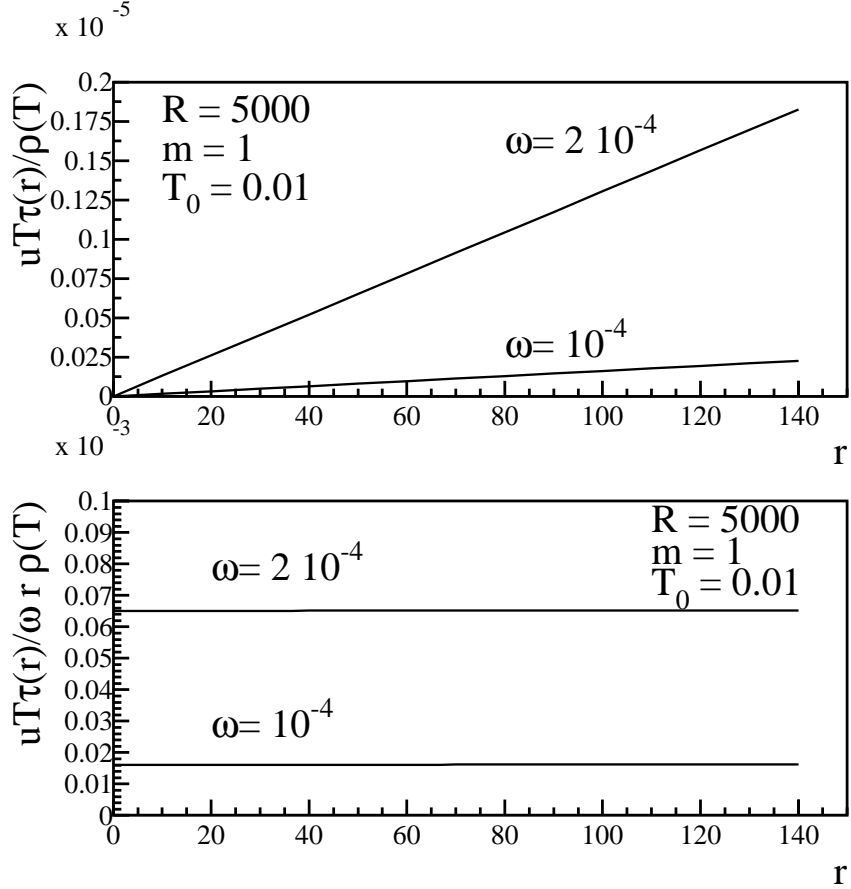}
\caption{Ratio between the projection $u_\mu \hat\tau_\nu T^{\mu\nu}$ for the free scalar 
field at global thermodynamical equilibrium within a rotating cylinder and the usual 
energy density of an ideal scalar gas as a function of the radial distance $r$. The radius $R$
is in arbitrary units and the values of $\omega$, $T_0$, $R$ and $m$ lie in the 
non-relativistic domain. 
\label{projection}}
\end{figure}
\end{center} 

As an example we calculate the stress-energy tensor of the free scalar real field 
(for details see Appendix A). The boundary condition at the outer surface $r=R$ of the 
cylinder is $\wpsi(R)=0$ which ensures the necessary vanishing of the flux eq.~(\ref{rhoflux}). 
Indeed, since $\wpsi(R)=0$, the gradient of the field at $r=R$  is normal to the 
outer surface, that is:
\be\label{boundary}
  \partial_\mu \wpsi |_{r=R} = \widehat \chi (t,R,\phi,z)\hat r^\mu.
\ee
Since for the free scalar field \footnote{The round brackets on indices stand for 
symmetrization.}
\be\label{setscal}
  \wT^{\mu\nu} = \partial^{(\mu} \wpsi \, \partial^{\nu)} \wpsi - g^{\mu\nu} 
  \widehat{\cal L}\qquad\qquad    
  \widehat{\cal L} = \frac{1}{2} \left( \partial_\mu \wpsi \, \partial^\mu \wpsi - 
  m^2 \wpsi^2 \right)
\ee
one has
$$
  \hat r_\mu \wT^{\mu\nu}(R) \beta_\nu = 0    
$$
hence the eq.~(\ref{rhoflux}), taking into account that $\xi=0$. Furthermore, the 
condition (\ref{boundary}) make, at the operatorial level, the fluxes of the energy 
and angular momentum outside the cylinder boundary vanishing, implying the conservation
of $\widehat H$ and $\widehat J_z$, as it should be. 

Unlike for the ideal case (\ref{tideal}), $T^{\mu\nu} \beta_\nu$ is not parallel 
to $\beta^\mu$. Indeed, one has
\be\label{uttau}
\hat\tau_\mu T^{\mu\nu} \beta_\nu = \sqrt{\beta^2}
2\gamma^2 \sum_{M=-\infty}^{+\infty}\sum_{p_T}
 \int \di p_L \; \frac{J_M^2(p_T r)}{(2\pi)^2 \, \varepsilon \, R^2 \, 
 J^{\prime 2}_M(p_T R)\frac{}{}} \frac{1}{{\rm e}^{(\varepsilon-M\omega)/T_0}-1} 
 \left[ \frac{}{} \omega r 
 \left( \varepsilon^2 + \frac{M^2}{r^2} \right) - (1 + \omega^2 r^2)
 \frac{\varepsilon M}{r} \right]
\ee
where $p_T$ are the discrete values related to the zeroes $\zeta_{l,M}$ of the Bessel 
function $J_M$ by $p_T R= \zeta_{l,M}$ and $\varepsilon = \sqrt{p_T^2 + p_L^2 + m^2}$, 
see Appendix A. As expected, the above expression is vanishing for $\omega=0$, i.e.
in the non-rotating case, but for $\omega \ne 0$ is non vanishing, what is confirmed 
by numerical computation shown in fig.~\ref{projection}. 
This computation was carried out in the non-relativistic limit with $m \gg T_0$ 
and for $r$ values such that $\omega r \ll 1$. Still, the covered range in $r$ far exceeds 
the typical microscopic length, that is the thermal wavelength scale $1/\sqrt{m T_0}$
which turns out to be 0.1 in the distance units of fig.~\ref{projection}. It can be 
seen that the ratio between the stress-energy tensor component $\tau \cdot T \cdot u$ 
in eq.~(\ref{uttau}) and the usual energy density expression of an ideal massive scalar 
is non-vanishing. It increases linearly as a function of $r$ which is owing to the 
fact that $\tau \cdot T \cdot u \simeq \tau^0 u^0 T^{00}$ in the non-relativistic limit 
and that $\tau^0 \simeq \omega r$ (see definition (\ref{tauvers})). Once $\tau \cdot T 
\cdot u$ is divided by $\omega r$, it can be seen that the ratio between this component 
and the usual energy density is approximately constant for small $r$. Moreover, its
value is of the order $(\omega/T_0)^2 = 10^{-4}$ with an exact scaling with $\omega^2$,
as expected. This demonstrates the difference between $\beta$ and Landau frame.

The full stress-energy tensor turns out to be
\begin{equation*}
 T = 
 \left( \begin{array}{cccc} 
                     u \cdot T \cdot u & u \cdot T \cdot \hat\tau & 0 & 0  \\
                     u \cdot T \cdot \hat\tau & \hat\tau \cdot T \cdot \hat\tau & 0 & 0 \\
		     0 & 0 & \hat r \cdot T \cdot \hat r & 0 \\
		     0 & 0 & 0 & k \cdot T \cdot k 
  \end{array} \right) 
\end{equation*}
where the quoted components are non-vanishing and calculated in Appendix A. In 
conclusion, $\hat \beta$ is not the timelike eigenvector $u_L$ of $T$ and the Landau 
and the $\beta$ frame are not equivalent in this case. While it is possible to 
express $u_L$ as superposition of $u = \hat\beta$ and $\tau$, the two spacelike 
eigenvectors $n$ and $k$ have different eigenvalues, because (see again Appendix A):
$$
 \hat r \cdot T \cdot \hat r - k \cdot T \cdot k = 
 \sum_{M=-\infty}^{+\infty}\sum_{p_T} \int \di p_L \; 
 \frac{2}{(2\pi)^2 \, \varepsilon \, R^2 \, 
 J^{\prime 2}_M(p_T R)\frac{}{}} \frac{1}{{\rm e}^{(\varepsilon-M\omega)/T_0}-1} 
 \left[ p_T^2 J^{\prime}_M(p_T r)^2 - p^2_L J_M(p_T r)^2 \right],
$$
which is not vanishing. This can be readily checked by setting $r=R$ and using the 
boundary condition of the Bessel function. Consequently, at thermodynamical 
equilibrium, the term $\Pi$ in the decomposition (\ref{landaudec}) is non-vanishing, 
unlike commonly assumed. Another important consequence of the difference between
$\beta$ and Landau frame is that the basic local thermodynamic relation (\ref{localtd2})
cannot be the same in both frames (see Appendix B).

\section{Equations of relativistic hydrodynamics in the $\beta$ frame}
\label{relhyd}

The $\beta$ frame is an especially suitable framework to write the equations of 
relativistic hydrodynamics. As it is well known, the general problem is to determine 
the evolution of the stress-energy tensor, and possibly several vector currents 
starting from definite initial conditions, under the assumption of approximate local 
thermodynamical equilibrium. This condition, in case of one conserved current, as 
we have seen, reduces the number of unknown functions to 5, that is the four components 
of $\beta$ and $\xi$, which is just the number of continuity equations. In terms of 
these variables, the equations of relativistic hydrodynamics do not show any distinction 
between equations of motion and equation of state (which is encoded in the dependence 
of the pressure on $\beta^2$ and $\xi$, as we will see). 

\subsection{Ideal hydrodynamics}

As we have seen in sect.~\ref{setensor} the stress-energy tensor in $x$ at the lowest order 
in the gradient expansion can be approximated by the ideal one $T^{\mu\nu}_{\rm id}$ 
with inverse temperature four-vector and the chemical potential equal to those in the point 
$x$. We have shown in ref.~\cite{becacov} that it can be obtained by taking derivatives 
of the thermodynamic potential current $\phi^\mu = p \beta^\mu$, where $p$ is the 
equilibrium pressure, a scalar field depending on the scalars $\beta^2$ and $\xi$. Thus
\be
  T^{\mu \nu}_{\rm id} = - 2 \frac{\partial p}{\partial \beta^2} 
  \beta^\mu \beta^\nu - p g^{\mu \nu},
\ee
being the derivative of the pressure proportional to the proper enthalpy density
\be\label{enth}
   - 2 \frac{\partial p}{\partial \beta^2} = \frac{\rho + p}{\beta^2} = 
   \frac{h}{\beta^2}.
\ee
Similarly
$$
  j^{\mu}_{\rm eq} = \frac{\partial p}{\partial \xi} \beta^\mu,
$$
being the derivative of the pressure with respect to $\xi$ proportional to
the charge density
$$
  \frac{\partial p}{\partial \xi} = \frac{q}{\sqrt{\beta^2}}.
$$
These expressions allow to reformulate ideal relativistic hydrodynamics through
5 unknown functions: the four-vector $\beta$ (whose modulus is the inverse local
temperature) and the scalar field $\xi$, corresponding to the ratio $\mu/T$. 
There are indeed 5 differential equations corresponding to the conservation equations 
of $T$ and $j$, which in principle allow to solve the problem, provided that the 
functional relation $p(\beta^2,\xi)$ is known, which is but the {\em complete} 
equation of state. Nothing new, however the introduction of these variables as primary 
fields allows to gain further insight into the structure and features of relativistic 
hydrodynamics.

At the lowest order in the gradients, using eq.~(\ref{setexpa}), the continuity equations
are those of the ideal hydrodynamics
\be\label{ideal1}
 \partial_\mu T^{\mu \nu}_{\rm id} = - 2 \frac{\partial^2 p}{\partial\beta^{22}}
 \beta^\mu \beta^\nu \partial_\mu \beta^2 - 2 \frac{\partial p}{\partial\beta^2\partial\xi}
 \beta^\nu \beta^\mu \partial_\mu \xi - 2 \frac{\partial p}{\partial\beta^2}
 ( \beta^\nu \partial \cdot \beta + \beta^\mu \partial_\mu \beta^\nu) - 
 \frac{\partial p}{\partial\beta^2} \partial^\nu \beta^2 - 
 \frac{\partial p}{\partial\xi} \partial^\nu \xi = {\cal O}(\partial^2) \simeq 0
\ee
and
\be\label{ideal2}
 \partial_\mu j^{\mu}_{\rm eq} =  \frac{\partial^2 p}{\partial\xi^2} 
 \beta^\mu \partial_\mu \xi + \frac{\partial^2 p}{\partial\beta^2\partial\xi} 
 \beta^\mu \partial_\mu \beta^2 +\frac{\partial p}{\partial \xi} \partial 
 \cdot \beta = {\cal O}(\partial^2) \simeq 0;
\ee
Being:
$$
   \partial \cdot \beta = \frac{D \beta^2}{2\sqrt{\beta^2}} + \nabla \cdot \beta
$$
the eq.~(\ref{ideal2}) can be written, at the lowest order, as
\be\label{ideal2new}
 \frac{\partial^2 p}{\partial\xi^2} \sqrt{\beta^2} D \xi + 
 \frac{\partial^2 p}{\partial \beta^2 \partial\xi} \sqrt{\beta^2} D \beta^2 
 +\frac{\partial p}{\partial \xi} \partial \cdot \beta = 0
\ee
and the eq.~(\ref{ideal1}), at the lowest order, can be split into two equations 
projecting along $\beta$ and transversely to it using \eqref{derdef}
\bea\label{ideal1new}
 && \frac{\partial^2 p}{\partial\beta^{22}} \beta^2 \sqrt{\beta^2} D \beta^2 
 + \frac{\partial^2 p}{\partial\beta^2\partial\xi} \beta^2 \sqrt{\beta^2}
 D \xi + \frac{\partial p}{\partial\beta^2} \beta^2 \left( \frac{3}{2 \sqrt{\beta^2}}
 D \beta^2 + \nabla \cdot \beta \right) + \frac{1}{2} \frac{\partial p}{\partial\xi} 
 \sqrt{\beta^2} D \xi = 0 \nonumber \\
 && \frac{\partial p}{\partial\beta^2} \left( \sqrt{\beta^2} 
 \Delta_{\mu\nu} D \beta^\nu + \frac{1}{2} \nabla_\mu \beta^2 \right)
 + \frac{1}{2} \frac{\partial p}{\partial\xi} \nabla_\mu \xi = 0.
\eea
These two equations are the relativistic generalizations of the continuity equations
and the Euler equation of motion of the fluid. We can readily retrieve its familiar
form by noting that
$$ 
 \frac{1}{2}  \frac{\partial p}{\partial\beta^2} \nabla_\mu \beta^2 
 + \frac{1}{2} \frac{\partial p}{\partial\xi} \nabla_\mu \xi = \frac{1}{2} 
 \nabla_\mu p
$$
and
$$
  \sqrt{\beta^2} \Delta_{\mu\nu} D \beta^\nu = \sqrt{\beta^2} \Delta_{\mu\nu} 
  D \left( \frac{1}{T} u^\nu \right) = \Delta_{\mu\nu} \beta^2 D u^\nu = 
  \beta^2 A_\mu,
$$
$A^\nu = D u^\nu$ being the acceleration by definition. The second of (\ref{ideal1new})
then becomes, by using (\ref{enth})
$$
 \frac{\partial p}{\partial\beta^2} \beta^2 A_\mu = -\frac{1}{2} (\rho + p) A_\mu
 = -\frac{1}{2} \nabla_\mu p
$$
that is the well known form of the relativistic Euler equation.

It is interesting to note that the first term in the relativistic Euler equation
in (\ref{ideal1new}) can also be written as
\be\label{flux1}
 \sqrt{\beta^2} \Delta_{\mu\nu} D \beta^\nu + \frac{1}{2} \nabla_\mu \beta^2
 = \beta_\lambda \Delta_{\mu\nu} ( \partial^\lambda \beta^\nu + 
   \partial^\nu \beta^\lambda )
\ee
as well as
\be\label{flux2}
\sqrt{\beta^2} \Delta_{\mu\nu} D \beta^\nu + \frac{1}{2} \nabla_\mu \beta^2
= \beta^2 A^\mu + \frac{1}{2} \nabla_\mu \frac{1}{T^2} = 
\frac{1}{T^2} \left( A_\mu - \frac{1}{T} \nabla_\mu T \right).
\ee
One can recognize in this expression the four-vector which the heat flow $q^\mu$
is proportional to in the first order dissipative hydrodynamics. Hence, we can 
say that the ideal relativistic Euler equation amount to state that the first-
order dissipative heat flow is parallel to the first-order dissipative current
proportional to $\nabla \xi$. For an uncharged fluid, it simply states that at
the first-order in the gradient expansion, this dissipative current vanishes.

We can use (\ref{ideal2new}) to obtain $D\xi$ as a function of $D\beta^2$ and
plug into the first equation of (\ref{ideal1new}), which then becomes
\bea
 && \left[ \frac{\partial}{\partial\beta^{2}} \big( (\beta^2)^{3/2} 
  \frac{\partial p}{\partial\beta^2} \big) -
  {(\frac{\partial^2 p}{\partial\beta^2\partial\xi})^2(\beta^2)^{3/2} \over 
  \frac{\partial^2 p}{\partial\xi^2}} -
  {\frac{\partial p}{\partial\xi}\frac{\partial^2 p}{\partial\beta^2\partial\xi} 
  \over \frac{\partial^2 p}{\partial\xi^2}}\sqrt{\beta^2}-\frac{(\frac{\partial p}
  {\partial \xi})^2}{4\sqrt{\beta^2}\frac{\partial ^2 p }{\partial \xi ^2}} 
  \right] D \beta^2 \nonumber \\
  && + \left[ \frac{\partial p}{\partial\beta^2} \beta^2- 
  \frac{\frac{\partial p}{\partial \xi}\left(\beta^2 
  \frac{\partial ^2 p}{\partial \beta^2\partial \xi}+
  \frac{1}{2}\frac{\partial p}{\partial \xi}\right)}{\frac{\partial^2 p}{\partial\xi^2}} 
  \right]\nabla \cdot \beta = 0.
\eea
This formula allows to obtain the derivative of $\beta^2$ along the flow as a
function of $\nabla \cdot \beta = (1/T) \nabla \cdot u$. Similarly, one can obtain 
the transverse gradient of $\xi$ as a function of the derivatives of 
$\beta$ through the (\ref{ideal1new}). It should be kept in mind that these relations
hold up to terms of the second order in the gradients. They can be used to eliminate
some of the gradients in the first-order expansion of the stress-energy tensor, 
or, better, to replace some of the first-order gradients with transverse gradients
of the $\beta$ field plus further corrections of the second order.

\subsection{Dissipative hydrodynamics in the $\beta$ frame}

Relativistic dissipative hydrodynamics has been the subject of intense investigations 
over the past decade \cite{Huovinen:2008te,Heinz:2005bw,Muronga:2001gn,Muronga:2003ta,
various1,various2,various3,various4,various5,various6} and an exhaustive 
discussion is well beyond the scope of this work. Herein, we confine ourselves to show 
that the $\beta$ frame is best suited to approach dissipative relativistic hydrodynamics 
as a gradient expansion. The main reason thereof has been mentioned in sect.~\ref{setensor}, 
that is the eqs.~(\ref{tevol}) expressing the density operator as a function
of the present "time" local equilibrium operator and an integral of the gradients of
the $\beta$ and $\xi$ fields. The expansion has been briefly outlined in sect.~\ref{setensor}
for the stress-energy tensor, but it can be extended to any observable. 

We are now going to show that indeed, in the familiar 1st order (Navier-Stokes) 
dissipative hydrodynamics, the transverse gradients of the velocity field and of the 
temperature can be re-expressed in terms of the gradients of $\beta$. We have already 
shown (see eqs.~(\ref{flux1},\ref{flux2}) that the combination
$$
 \left( A_\mu - \frac{1}{T} \nabla_\mu T \right)
$$
appearing in the familiar form of the heat current vector can be rewritten in a compact 
way in terms of the gradients of $\beta$. Similarly, the transverse gradients of the 
velocity field $\nabla^\mu u^\nu$ can be written as follows
\bea
  \nabla_\mu u^\nu && = \nabla_\mu \frac{\beta^\nu}{\sqrt{\beta^2}} = \beta^\nu
 \left( -\frac{1}{2}\right) (\beta^2)^{-3/2} \nabla_\mu \beta^2 + \frac{1}{\sqrt{\beta^2}}
 \nabla_\mu \beta^\nu \nonumber \\
 && = \frac{1}{\sqrt{\beta^2}} \left( - \frac{\beta^\nu \beta^\rho}{\beta^2} 
 \nabla_\mu \beta_\rho + \nabla_\mu \beta^\nu \right) = 
 \frac{1}{\sqrt{\beta^2}} \Delta^{\rho\nu} \nabla_\mu \beta_\rho,
\eea
where we have used the definition (\ref{derdef}). Hence, the Navier-Stokes shear term
can be fully expressed in terms of the inverse temperature four-vector. Likewise, it 
is easy to show that the expansion term
$$
   \nabla_\mu u^\mu = \frac{1}{\sqrt{\beta^2}} \nabla_\mu \beta^\mu.
$$
%

\section{Conclusions}
\label{conclu}

We conclude with a short recapitulation of the main findings of this work:
\begin{itemize}
\item{} The notion of relativistic local thermodynamical equilibrium (LTE) can be 
defined independently of kinetic theory, in a form which is suitable for a strongly
interacting fluid. 
\item{} Local thermodynamical equilibrium notion is, by construction, frame dependent. 
There is a preferred frame for it, the one where basic thermodynamics relations 
take on the simplest form, what we have called the $\beta$ frame; the $\beta$ 
frame is {\em the} frame when expansions from LTE are to be carried out.
\item{} Physically, the $\beta$ four-vector direction is identified by the four-velocity 
of an idealized relativistic thermometer at equilibrium with the system.
\item{} The $\beta$ frame has many interesting features in relativistic hydrodynamics,
both ideal and dissipative. The $\beta$ four-vector and the other intensive parameter
$\xi = \mu/T$ are the solutions of the eqs.~(\ref{betaeq2}) for a non-vorticous $\beta$
field, or, in general, of the eqs.~(\ref{betaeq3}).
\item{} The $\beta$ frame in general differ from both Eckart and Landau frames.
It differs from those frames in situations where a local acceleration
is present, like in the rotating fluid. The local acceleration or rotation provides
a new independent macroscopic scale which introduces quadratic corrections to
the stress-energy tensor.
\end{itemize}

Furthermore, we have seen that the familiar ideal hydrodynamic equations of motion
can be written in a form where $\beta$ and $\xi$ are the 5 unknown fields. Also, 
first order dissipative hydrodynamics can be written in a form where the gradients 
are, again, only those of $\beta$ and $\xi$. It would be very interesting to extend 
the Israel-Stewart theory of causal hydrodynamics in terms of these fields and assess 
the stability of the equations.

\section*{Acknowledgments}   

L.T. acknowledges financial support from the Polish National Science Center 
grant No. DEC-2012/06/A/ST2/00390.



\section*{APPENDIX A - Stress-energy tensor for the free scalar field}

The Klein-Gordon equation of the real scalar field in cylindrical coordinates with 
Dirichlet boundary conditions $\wpsi(R)=0$ has the eigenfunctions:
\begin{equation}\label{particles}
 f_{\bf n} = C_{\bf n} J_M(p_T r) 
 \exp\left[ -i\left(\frac{}{} \varepsilon_{\bf n} t - p_L z - M \phi \right) \right],
\end{equation}
where $p_L$ is a continuous longitudinal momentum, $M$ is the integer angular momentum 
quantum number and the (discrete) transverse momenta $p_T(M,l)$ with $l=0,1,\ldots$ 
are the solutions of the boundary condition equation:
\begin{equation}\label{Bessel}
 J_M(p_T R)=0
\end{equation}
In the above two equations, $J_M$ is the Bessel function of integer order $M$. The
${\bf n} = (p_L,M,l(M))$ is the vector of quantum numbers and the energy $\varepsilon_{\bf n}$ 
and the normalization coefficient $C_{\bf n}$ read:
\begin{equation}\label{cn}
 \varepsilon_{\bf n}=\sqrt{m^2+p_z^2+p_{T}^2} \qquad \qquad
  C_{\bf n}^2 = \frac{1}{(2\pi)^2 \varepsilon_{\bf n} R^2 J^{\prime 2}_M(p_T R)}.
\end{equation}
The eigenfunctions $f_{\bf n}$ are orthogonal:
\begin{eqnarray*}
 && \int \di^3 \x \; f^*_{\bf n}f_{\bf n^\prime} =  
 C_{\bf n}C_{\bf n^\prime} \int \di^3 \x \; J_M(p_T r)J_{M^\prime}(p_T^\prime r) 
 \exp\left\{ i\left[ \left(\frac{}{} \varepsilon_{\bf n} - \varepsilon_{\bf n^\prime} 
 \right) t - \left( p_L - p_L^\prime \right) z - \left( M -M^\prime \right) \phi \right] 
 \right\} \nonumber \\ 
 && = C_{\bf n}C_{\bf n^\prime}(2\pi)^2 \delta(p_L-p_L^\prime)\delta_{M,M^\prime} 
 \int_0^R \!\!\!\! {\rm d} r \; r  J_M(p_T r)  J_M(p_T^\prime r)
  \exp\left[ i\left(\frac{}{} \varepsilon_{\bf n} - \varepsilon_{\bf n^\prime} \right) 
  t \right] = C_{\bf n}^2 (2\pi)^2 \frac{R^2}{2}J^\prime_M(p_T R)^2 \, \delta_{\bf n, n^\prime} ,
\end{eqnarray*}
where in the last equality we took advantage of the orthogonality relations of Bessel 
functions and:
\be\label{delta}
   \delta_{\bf n, n^\prime} \equiv \delta(p_L-p'_L) \delta_{MM'} \delta_{ll'}
\ee
The full orthogonality relations can be rewritten in the more compact form by using
the normalization in eq.~(\ref{cn}):
\begin{equation}\label{ort}
 \int \di^3 \x \; f^*_{\bf n}f_{\bf n^\prime} = \frac{1}{2\varepsilon_{\bf n}} 
 \delta_{\bf n, n^\prime}.
\end{equation}
Another useful relation is:
\begin{eqnarray}\label{star}
  \int \di^3 \x \; f^*_{\bf n}f^*_{\bf n^\prime} &=&  C_{\bf n} 
  C_{\bf n^\prime} \int \di^3 \x \;  J_M(p_T r)J_{M^\prime}(p_T^\prime r) 
  \exp\left\{ i\left[ \left(\frac{}{} \varepsilon_{\bf n} + \varepsilon_{\bf n^\prime} 
  \right) t - \left( p_L + p_L^\prime \right) 
  z - \left( M + M^\prime \right) \phi \right] \right\} \\ \nonumber
  &=& C_{\bf n}C_{\bf n^\prime} (2\pi)^2 \delta(p_L+p_L^\prime)\delta_{M,-M^\prime}  
  \int_0^R \!\!\!\! {\rm d} r \; r  J_M(p_T r)  J_{M^\prime}(p_T^\prime r) 
  \exp\left[ i\left(\frac{}{} \varepsilon_{\bf n} + \varepsilon_{\bf n^\prime} \right) 
  t \right] \\ \nonumber
  &=& \frac{1}{2\varepsilon_{\bf n}}(-1)^M \exp\left( 2i \frac{}{} \varepsilon_{\bf n} 
  \, t \right) \delta_{\bf n^\prime,\tilde n},
\end{eqnarray}
where ${\bf \tilde n}=(-p_L,-M, l)$\footnote{It is important to note how 
$\varepsilon_{\bf \tilde n}=\varepsilon_{\bf n}$ and $C_{\bf \tilde n}=C_{\bf n}$ } 
and in the last equality we used the $J_{-M}=(-1)^M J_M$ relation among the integer 
Bessel funtions.

The field operator reads:
\begin{equation}\label{x}
 \wpsi (x) = \sum_{\bf n}\left[ \frac{}{} f_{\bf n} \, a _{\bf n} + f^*_{\bf n} \, 
 a^\dagger_{\bf n} \right].
\end{equation}
From eqs.~(\ref{ort}), (\ref{star}) and the canonical equal time commutation relations:
\begin{equation}\label{canonical}
 [ \wpsi(t,{\bf x}), \wpsi(t,{\bf y}) ] = [\partial_t \wpsi(t,{\bf x})  
 \equiv {\widehat \Pi}(t,{\bf x}) , 
 {\widehat \Pi}(t,{\bf y}) ] = 0 \qquad \quad [ \wpsi(t,{\bf x}), 
 {\widehat \Pi}(t,{\bf y})] = i \delta^3({\bf x} - {\bf y}),
\end{equation}
the commutation relations between creation and annihilation operators 
$a^\dagger_{\bf n}$, $a_{\bf n}$ follow:
\be\label{aa}
 [a_{\bf n}, a_{\bf n^\prime}] = 0 \qquad \qquad 
 [a^\dagger_{\bf n}, a^\dagger_{\bf n^\prime}]=0
 \qquad  [a_{\bf n}, a^\dagger_{\bf n^\prime}] = \delta_{\bf n, n^\prime}
\ee

Reasoning as in ref.~\cite{becatinti1}, one can readily show that:
\begin{equation}\label{Bose}
 \langle a_{\bf n}a_{\bf n^\prime}\rangle = \langle a^\dagger_{\bf n} 
 a^\dagger_{\bf n^\prime}\rangle = 0 
 \qquad \quad \langle a^\dagger_{\bf n}a_{\bf n^\prime}\rangle  = 
 {\frac{1}{{\rm e}^{(\varepsilon_{\bf n} -M\omega)/T_0} -1 }}\delta_{\bf n,n^\prime},
\end{equation}
where in the last term one can recognize the typical Bose statistics mean occupation
number henceforth denoted as $n_B$:
$$
  n_B \equiv 
  {\frac{1}{{\rm e}^{(\varepsilon_{\bf n} -M\omega)/T_0} -1 }}
$$
We can now calculate the projections of the mean stress-energy tensor $\langle 
: \wT : \rangle$ with $\wT$ like in eq.~(\ref{setscal}), in the base $\{u,n,k,\tau\}$ .
First, we calculate the mean value $\langle:\widehat{\cal L}:\rangle$ of the Lagrangian 
density in ref.~(\ref{setscal}); for this purpose, one needs derivatives of the field:
\begin{eqnarray*}
 &&  \partial_r \wpsi = \sum_{\bf n}\left[\frac{}{} 
 (\partial_r f_{\bf n})a_{\bf n} + (\partial_r f^*_{\bf n})a^\dagger_{\bf n} \right] 
 \qquad \qquad \partial_z \wpsi = \sum_{\bf n}(ip_L)\left[\frac{}{}  
 f_{\bf n}a_{\bf n} - f^*_{\bf n}a^\dagger_{\bf n} \right] \\ 
 && \partial_\varphi \wpsi = \sum_{\bf n}(iM)\left[\frac{}{}  
 f_{\bf n}a_{\bf n} - f^*_{\bf n}a^\dagger_{\bf n} \right]. \label{r_phi_z}
\end{eqnarray*}
Dialing the above expansions in the lagrangian in eq.~(\ref{setscal}), one obtains:
\begin{eqnarray*}
 \langle :\widehat {\cal L}:\rangle &=& 
 \frac{1}{2} \langle:(\partial_t\wpsi)^2  - (\partial_x\wpsi)^2 - (\partial_y\wpsi)^2 
 - (\partial_z\wpsi)^2 - m^2 \wpsi^2:\rangle = \frac{1}{2}\langle :(\partial_t\wpsi)^2  
 - (\partial_r\wpsi)^2 - \frac{1}{r^2}(\partial_\varphi\wpsi)^2 - (\partial_z\wpsi)^2 - 
 m^2 \wpsi^2:\rangle \\ \nonumber
 &=& \sum_{\bf n} \bn\left\{ \varepsilon_{\bf n}^2 |f_{\bf n}|^2 - 
 |\partial_rf_{\bf n}|^2 -\frac{M^2}{r^2} |f_{\bf n}|^2 -p_L^2 |f_{\bf n}|^2 -m^2  
 |f_{\bf n}|^2 \right\} = 
 \sum_{\bf n} \bn \left\{ \left( p_T^2 - \frac{M^2}{r^2} \right) |f_{\bf n}|^2 
 -  |\partial_r f_{\bf n}|^2 \right\}.
\end{eqnarray*}

Every projection involving one $k$ is vanishing as $k\cdot\partial=\partial_z$  
involves a multiplication of each term within the sum $\sum_{\bf n}$ by $p_L$. 
On the other hand, the $k \cdot T \cdot k$ diagonal term reads:
$$
  k\cdot T\cdot k = \langle :(\partial_z \wpsi)^2 + {\cal L}:\rangle = \sum_{\bf n}\bn 
  \left\{ 2 \, p_L^2\,  |f_{\bf n}|^2 + \left( p_T^2 - \frac{M^2}{r^2} \right) 
  |f_{\bf n}|^2 -  |\partial_r f_{\bf n}|^2 \right\}.
$$
Similarly, for the projections along $n$, the off-diagonal $\hat r \cdot T \cdot \hat\tau$
and $\hat r \cdot T \cdot u$ terms vanish because:
\begin{eqnarray*}
&& \langle :\partial_{(t}\wpsi\partial_{r)}\wpsi:\rangle = \sum_{\bf n} \bn 
 \left\{ \frac{}{} -i\varepsilon_{\bf n}  f^*_{\bf n}(\partial_r f_{\bf n}) + 
 i\varepsilon_{\bf n}  f^*_{\bf n}(\partial_r f_{\bf n})  \right\} = 0  \nonumber \\
&& \langle :\partial_{(\varphi}\wpsi\partial_{r)}\wpsi:\rangle = \sum_{\bf n} \bn 
 \left\{ \frac{}{} iM  f_{\bf n}(\partial_r f^*_{\bf n}) - 
 iM  f^*_{\bf n}(\partial_r f_{\bf n})  \right\} = 0,
\end{eqnarray*}
taking into account that $f_{\bf n}(\partial_r f^*_{\bf n})$ is real. On the other
hand:
$$
 \hat r \cdot T \cdot \hat r = \langle :(\partial_r \wpsi)^2 + \widehat{\cal L}:\rangle = 
 \sum_{\bf n}\bn \left[ |\partial_r f_{\bf n}|^2 + \left( p_T^2 - 
 \frac{M^2}{r^2} \right) |f_{\bf n}|^2 \right].
$$
Using the relations:
$$
 u\cdot\partial = \gamma \, \partial_t +\gamma \, v 
 \frac{1}{r}\partial_\varphi = \gamma \, \partial_t +\gamma\, \omega \, 
 \partial_\varphi \qquad \qquad \hat\tau \cdot\partial = \gamma \, v \, 
 \partial_t +\gamma\frac{1}{r}\partial_\varphi,
$$
we can calculate the diagonal projections onto $u$ and $\hat\tau$:
\begin{eqnarray} \nonumber
 u\cdot T \cdot u &=& \langle :\gamma^2\left[ ( \frac{}{}\partial_t \wpsi)^2 
 + \omega^2 ( \frac{}{}\partial_\varphi \wpsi)^2 + 2\omega( \frac{}{}\partial_{(t} \wpsi 
 \partial_{\varphi)} \wpsi) \right] -{\cal L} :\rangle = \\ \nonumber
 &=& \sum_{\bf n} \bn \left\{ 2\gamma^2\left[ \frac{}{}\varepsilon_{\bf n}^2 + M^2\omega^2 
 - 2\varepsilon_{\bf n}M\omega \right] |f_{\bf n}|^2 - \left( p_T^2 - \frac{M^2}{r^2} 
 \right) |f_{\bf n}|^2 +  |\partial_r f_{\bf n}|^2 \right\} = \\ \label{varepsilon}
 &=&  \sum_{\bf n} \bn \left\{ 2\gamma^2\left[ \frac{}{} \varepsilon_{\bf n} - M \omega 
 \right] ^2 |f_{\bf n}|^2 - \left( p_T^2 - \frac{M^2}{r^2} \right) |f_{\bf n}|^2 + 
 |\partial_r f_{\bf n}|^2 \right\} \\ \nonumber \\ \nonumber \\ \nonumber
 \hat\tau\cdot T \cdot \hat\tau&=&  \langle :\gamma^2\left[ v^2( \frac{}{}\partial_t \wpsi)^2 
 +  ( \frac{1}{r}\partial_\varphi \wpsi)^2 + 2\omega( \frac{}{}\partial_{(t} \wpsi 
 \partial_{\varphi)} \wpsi) \right] + {\cal L} :\rangle = \\ \nonumber
 &=& \sum_{\bf n} \bn \left\{ 2\gamma^2\left[ \varepsilon_{\bf n}^2 \, \omega^2 \, r^2  
 +  \frac{M^2}{r^2} - 2\varepsilon_{\bf n}M\omega \right] |f_{\bf n}|^2 + 
 \left( p_T^2 - \frac{M^2}{r^2} \right) |f_{\bf n}|^2 -  |\partial_r f_{\bf n}|^2 \right\} 
 = \\ \label{ptau}
 &=&  \sum_{\bf n} \bn \left\{ 2\gamma^2\left[ \frac{}{} \varepsilon_{\bf n} \, v - 
 \frac{M}{r} \right] ^2 |f_{\bf n}|^2 + \left( p_T^2 - \frac{M^2}{r^2} \right) |f_{\bf n}|^2 
 - |\partial_r f_{\bf n}|^2 \right\}.
\end{eqnarray}
as well as the non-diagonal term:
\begin{eqnarray*}
 u\cdot T \cdot \hat\tau &=& \langle :\gamma^2 \left\{ v\left[ \left( \frac{}{} 
 \partial_t {\wpsi} \right)^2 + \frac{1}{r^2}\left( \frac{}{} \partial_\varphi {\wpsi} 
 \right)^2 \right] + \frac{1}{r}\left(1 + v^2 \right)\partial_{(t}{\wpsi} \, 
 \partial_{\varphi)}{\wpsi}\right\} :\rangle = \\ \label{nondiag_1}
 &=&  2\gamma^2 \sum_{\bf n} \bn \left\{ \omega r \left( \varepsilon_{\bf n}^2 + 
 \frac{M^2}{r^2} \right) - (1 + \omega^2 r^2)\frac{\varepsilon_{\bf n}M}{r} \right\} 
 |f_{\bf n}|^2,
\end{eqnarray*}
or, writing explicitly the $ |f_{\bf n}|^2$ function
\begin{equation}\label{nondiag_2}
 u \cdot T \cdot \hat\tau = 2\gamma^2 \sum_{\bf n} 
 \frac{J_M^2(p_T r)}{ (2\pi)^2 \, \varepsilon_{\bf n} \, R^2 \, J^{\prime 2}_M(p_T R)\frac{}{}} 
 \frac{1}{{\rm e}^{(\varepsilon_{\bf n}- M\omega)/T_0} -1 } \left[ \frac{}{} \omega r 
 \left( \varepsilon_{\bf n}^2 + \frac{M^2}{r^2} \right) - (1 + \omega^2 r^2)
 \frac{\varepsilon_{\bf n}M}{r} \right].
\end{equation}
%

\section*{APPENDIX B - Thermodynamic relations and change of frame}

In this section we show how the basic thermodynamic relation (\ref{basic}) between 
proper entropy density, proper energy and charge density, pressure and temperature
is modified by a change of frame. We first consider the familiar global equilibrium
case, where $\beta$ is a constant four-vector. It should be first pointed out that
in the case of global equilibrium with constant $\beta$, its direction coincides 
with both Landau and Eckart four-velocities, that is there exists {\em one} four-
vector to which all physical vectors are parallel \cite{israel}. According to our 
reasoning in sect.~\ref{local}, in this case the entropy current $s^\mu$ reads, as
it is known \cite{israel}:
$$
s^\mu = p \beta^\mu + T^{\mu\nu} \beta_\nu - \xi j^\mu.
$$
Note that $T^{\mu\nu}=T^{\mu\nu}_{\rm LE}$ and $j^\mu = j^\mu_{\rm LE}$ being at global
equilibrium; also note that the extra term $s^\mu_T$ in eq.~(\ref{current}) is now missing because the 
total entropy is conserved, thus it should be the same for any three-dimensional 
spacelike hypersurface chosen to integrate the entropy current. Let us now project 
onto an arbitrary frame $u$ the above expression, replacing first $T^{\mu\nu}$ with 
its expression at global equilibrium $(\rho+p) \hat\beta^\mu \hat\beta^\nu - p 
g^{\mu\nu}$ and likewise for $j^\mu = q \hat \beta^\mu$: 
\be\label{entframe}
s^\mu u_\mu \equiv s_u = p (\beta \cdot u) + \rho (\beta \cdot u) - 
\frac{\xi}{\sqrt{\beta^2}} q (\beta \cdot u).
\ee
The functions $p(\beta^2,\xi)$, $\rho(\beta^2,\xi)$ and $q(\beta^2,\xi)$ are, by 
definition, the usual thermodynamic functions, with $\beta^2 \equiv 1/T^2$ and $\xi
=\mu/T$. Now we remind that $\beta \cdot u$ is the inverse temperature marked by the 
thermometer moving at speed $u$, say $T_u$, hence the eq.~(\ref{entframe}) can be 
rewritten as:
\be
 T_u s_u = p + \rho - \mu q.
\ee
However, the functions $p$, $\rho$, $q$ are not formally the same thermodynamic functions 
of $T$ as of $T_u$. The difference between the two can be obtained by setting $\hat
\beta = u + \delta u$, whence:
$$
  \delta u \cdot u = -(\delta u)^2
$$
so that:
$$
 \frac{1}{T_u} = \beta \cdot u = \sqrt{\beta^2} (\hat \beta \cdot u) = \frac{1}{T}
 [1 - (\delta u)^2].
$$
Thus, the difference between the thermodynamic relation in the two frames is of the
second order in the difference between velocity fields.

Of course, in the global equilibrium case with constant $\beta$, as has been mentioned,
there is no difference between different frames. However, in LTE, the choice of the
frame changes the basic thermodynamic relation by terms of the second order in the
difference between velocities. 

In the case of global equilibrium with rotation, where, as we have shown, the $\beta$ 
and the Landau frames do not coincide, the basic thermodynamic relation cannot be 
the simplest one (\ref{localtd2}) in both frames. If $\phi^\mu = p \beta^\mu$ as in the 
previous case, then the $\beta$ frame is the one where (\ref{localtd2}) holds.

\end{document}